\documentclass[epsf]{emulateapj}
\usepackage{natbib}
\usepackage{subfigure}
\newcommand{\gps}{\ensuremath{g_{\rm P1}}}
\newcommand{\rps}{\ensuremath{r_{\rm P1}}}
\newcommand{\ips}{\ensuremath{i_{\rm P1}}}
\newcommand{\zps}{\ensuremath{z_{\rm P1}}}
\newcommand{\yps}{\ensuremath{y_{\rm P1}}}

\begin{document}
\submitted{Submitted to the Astrophysical Journal}

\title{Pan-STARRS1 Discovery of Two Ultra-Luminous Supernovae at z $\approx$ 0.9}

\author{L.~Chomiuk\altaffilmark{1, 2}, R.~Chornock\altaffilmark{2}, A.~M.~Soderberg\altaffilmark{2}, E.~Berger\altaffilmark{2}, M.~E.~Huber\altaffilmark{3}, G.~Narayan\altaffilmark{4}, A.~Rest\altaffilmark{5}, R.~A.~Chevalier\altaffilmark{6}, R.~J.~Foley\altaffilmark{2}, S.~Gezari\altaffilmark{3}, R.~P.~Kirshner\altaffilmark{2}, A.~Riess\altaffilmark{3}, S.~A.~Rodney\altaffilmark{3}, S.~J.~Smartt\altaffilmark{7}, C.~W.~Stubbs\altaffilmark{4}, J.~L.~Tonry\altaffilmark{8}, W.~M.~Wood-Vasey\altaffilmark{9}, W.~S.~Burgett\altaffilmark{8}, K.~C.~Chambers\altaffilmark{8}, H.~Flewelling\altaffilmark{8}, K.~Forster\altaffilmark{10}, N.~Kaiser\altaffilmark{8}, R.-P.~Kudritzki\altaffilmark{8}, D.~C.~Martin\altaffilmark{10}, J.~S.~Morgan\altaffilmark{8}, J.~D.~Neill\altaffilmark{10}, P.~A.~Price\altaffilmark{11}, and R.~J.~Wainscoat\altaffilmark{8}}
\email{lchomiuk@cfa.harvard.edu}
\altaffiltext{1}{National Radio Astronomy Observatory, P.O. Box O, Socorro, NM 87801 USA}
\altaffiltext{2}{Harvard-Smithsonian Center for Astrophysics, 60 Garden Street, Cambridge, MA 02138, USA}
\altaffiltext{3}{Department of Physics and Astronomy, Johns Hopkins University, 3400 North Charles Street, Baltimore, MD 21218, USA}
\altaffiltext{4}{Department of Physics, Harvard University, Cambridge, MA 02138, USA}
\altaffiltext{5}{Space Telescope Science Institute, 3700 San Martin Drive, Baltimore, MD 21218, USA}
\altaffiltext{6}{Department of Astronomy, University of Virginia, P.O. Box 400325, Charlottesville, VA 22904-4325}
\altaffiltext{7}{Astrophysics Research Centre, School of Mathematics and Physics, Queen's University Belfast, Belfast, BT7 1NN, UK}
\altaffiltext{8}{Institute for Astronomy, University of Hawaii at Manoa, Honolulu, HI 96822, USA}
\altaffiltext{9}{Department of Physics and Astronomy, University of Pittsburgh, 3941 O'Hara Street, Pittsburgh, PA 15260, USA}
\altaffiltext{10}{California Institute of Technology, 1200 E. California Blvd., Pasadena, CA  91125, USA}
\altaffiltext{11}{Department of Astrophysical Sciences, Princeton University, Princeton, NJ 08544, USA}

\begin{abstract}
We present the discovery and analysis of two ultra-luminous supernovae (SNe) at z $\approx0.9$ with the Pan-STARRS1 Medium-Deep Survey. These SNe, PS1-10ky and PS1-10awh, are amongst the most luminous SNe ever discovered, comparable to the unusual transient SCP 06F6. Like SCP 06F6, they show characteristic high luminosities ($M_{\rm bol} \approx -22.5$ mag), blue spectra with a few broad absorption lines, and no evidence for H or He. We have constructed a full multi-color light curve sensitive to the peak of the spectral energy distribution in the rest-frame ultraviolet, and we have obtained time-series spectroscopy for these SNe. Given the similarities between the SNe, we combine their light curves to estimate a total radiated energy over the course of explosion of $(0.9-1.4) \times 10^{51}$ erg. We find expansion velocities of $12,000-18,000$ km s$^{-1}$ with no evidence for deceleration measured $\sim$3 rest-frame weeks either side of light-curve peak, consistent with the expansion of an optically-thick massive shell of material. We show that radioactive decay is not sufficient to power PS1-10ky, and discuss two plausible origins for these events: the initial spin-down of a newborn magnetar in a core-collapse SN, or SN shock breakout from the dense circumstellar wind surrounding a Wolf-Rayet star.
\end{abstract}


\section{Introduction}

The observational and physical parameter space occupied by supernovae (SNe) has expanded dramatically because of the recent discovery of several ultra-luminous SNe. These SNe are significantly more luminous than Type Ia explosions, displaying absolute bolometric magnitudes at light curve maximum of $< -21$ mag and total radiated energies on the order of $10^{51}$ erg. Thus far, ultra-luminous SNe have displayed impressive diversity, ranging from the Type Ic SN 2007bi, proposed to be a pair-instability explosion \citep{GalYam_etal09, Young_etal10}, to the Type IIn SN 2006gy, with strong signs of circumstellar interaction \citep{Smith_etal07, Smith_McCray07, Ofek_etal07}. The mysterious transient SCP 06F6 discovered by \citet{Barbary_etal09} showed a largely featureless spectrum with a few broad absorption lines, and proved so perplexing that no redshift could be identified. Because it was uncertain if the source was Galactic or at a cosmological distance, many ideas were presented to explain it ranging from an outburst on a white dwarf to a broad-lined QSO \citep{Barbary_etal09, Gansicke_etal09, Chatzopoulos_etal09, Soker_etal10}.

Since SCP 06F6 was discovered by \citet{Barbary_etal09}, similar objects have been found with significant frequency in wide-field transient surveys. \citet{Quimby_etal11} collected five additional SCP 06F6-like sources discovered by the Texas Supernova Survey and the Palomar Transient Factory (PTF), including the earlier-identified SN 2005ap \citep{Quimby_etal07}. They identifed narrow \ion{Mg}{2} $\lambda\lambda$2796,2803 absorption lines in their spectra, presumably associated with the interstellar media of the host galaxies. From this absorption, they derived redshifts ranging from $0.25 \lesssim \textrm{z} \lesssim 0.5$ for their sources, and a redshift of z $= 1.19$ for SCP 06F6 itself. These cosmological distances imply that the SCP 06F6-like sources are some of the most luminous SNe known, with peak absolute magnitudes $M_{U} \lesssim -22$ mag and total radiated energies $\gtrsim 10^{51}$ erg. 

\begin{figure}
\centering
\includegraphics[width=8.5cm]{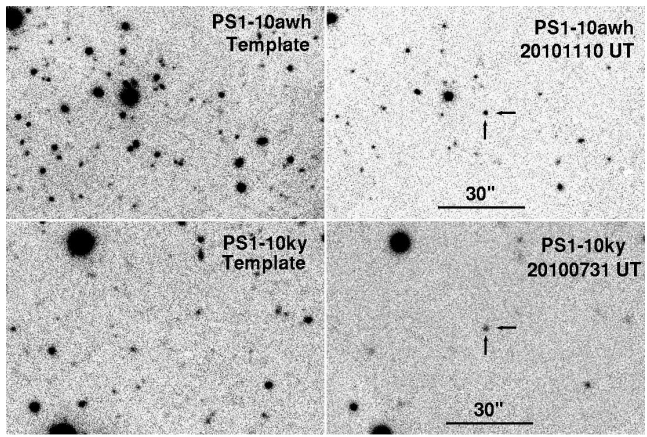}
\caption{Cut-outs of PS1 \ips-band images showing the region around PS1-10awh (top) and PS1-10ky (bottom). The left column shows stacked images using data before explosion; the right column shows images from single nights around maximum light.}
\label{comparison}
\end{figure}

With their redshifts known, a common set of observational properties began to emerge for the SCP 06F6-like objects. In addition to their very high peak luminosities, these sources all show distinctive symmetric light curves with rise and decline times on the order of 30 days in the rest frame. They also all show blue spectra with only a handful of features; \citet{Quimby_etal11} identify the broad absorption lines with light metals (C, O, Si, and Mg). There is no clear evidence for H or He in the spectra of an SCP 06F6-like transient, although \citet{Quimby_etal07} note a broad feature in SN 2005ap that may be associated with H$\alpha$. There also is no detection of strong \ion{Fe}{2} or \ion{Fe}{3} lines in the early or peak spectra of these transients. 

Subsequently, the ultra-luminous SN 2010gx was discovered by both the Catalina Real-time Transient Survey \citep{Mahabal_etal10, Mahabal_Drake10} and PTF \citep{Quimby_etal10}, and was determined to be at z = 0.23. \citet{Pastorello_etal10} carried out a detailed study of this source, which shows SCP 06F6-like spectral features and high luminosity at early times. They find that, a few weeks after peak light, this source begins to show iron and other features indicative of Type Ic SNe. They assert that SCP 06F6-like objects are likely to be associated with SNe Ibc, consistent with the lack of H in their spectra.

Two potential explanations for these SCP 06F6-like sources have surfaced: the interaction of the SN shock with a dense circumstellar shell of H-poor material \citep[e.g.,][]{Chevalier_Irwin11} or the spin-down of a new-born magnetar embedded in the SN ejecta \citep{Kasen_Bildsten10, Woosley10}. Both of these models have the potential to explain the high luminosities and association with SNe Ibc.

Here, we present two new ultra-luminous SNe at z $\approx$ 0.9 discovered with Panoramic Survey Telescope \& Rapid Response System 1 (Pan-STARRS1, abbreviated as PS1 here). By combining data for these two sources, we obtain the first full composite multi-color light curve for SCP 06F6-like objects and place constraints on the origins of these sources. Our photometry in multiple filters allows us to measure the color evolution of these sources, while their high redshifts aid in measuring bolometric luminosities, as these sources' spectral energy distributions (SEDs) peak in the rest-frame ultraviolet. Our spectroscopic observations sample the entire light curve, enabling us to measure the evolution of photospheric velocities. These measurements place important constraints on any model that attempts to explain these sources' high luminosities.

In \S \ref{obs}, we describe our data, obtained using PS1, MMT, Gemini, \emph{GALEX}, and the EVLA. In \S \ref{host} we constrain the properties of the host galaxies. In \S \ref{temp}, \ref{lum}, and \ref{vel}, we measure the evolution of their photospheric temperatures, bolometric luminosities, and photospheric velocities, combining our new data with published data of SCP 06F6 and SN 2010gx. In \S \ref{model}, we consider three possible physical scenarios that might explain the characteristics of these sources: radioactive decay, magnetar spin-down, and circumstellar interaction. Finally in \S \ref{conclude}, we summarize our results.

\section{Observations}\label{obs}

\subsection{PS1 Photometry} \label{phot}

\begin{figure}
\centering
\includegraphics[angle=90, width=9.5cm]{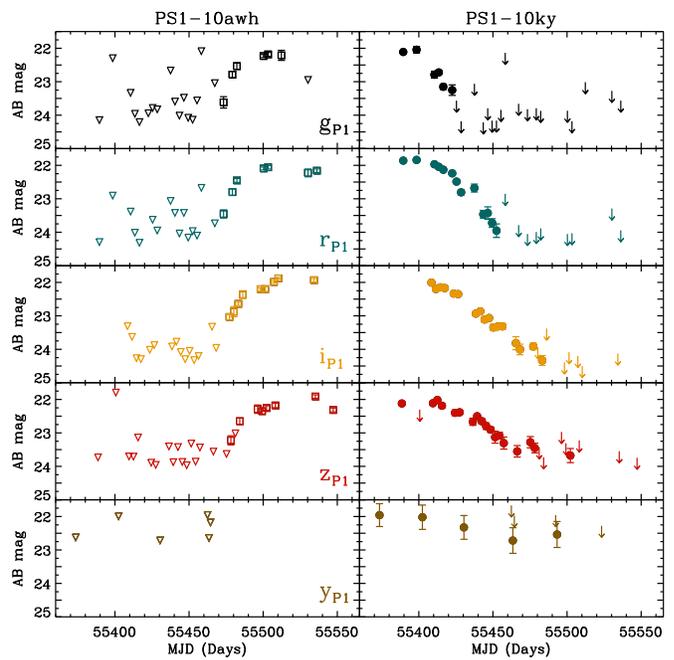}
\caption{Observed light curves for PS1-10awh (left column) and PS1-10ky (right column) in five filters from top to bottom: \gps\ (black), \rps\ (blue), \ips\ (gold), \zps\ (red), and \yps\ (brown). Apparent AB magnitudes are plotted as a function of modified Julian day, with squares and circles representing PS1-10awh and PS1-10ky respectively, and 3$\sigma$ upper limits denoted by triangles and arrows.}
\label{lc_app}
\end{figure}

\subsubsection{PS1 Survey Summary}
The PS1 system is a high-etendue wide-field imaging system, designed for dedicated survey observations. The system is installed on the peak of Haleakala on the island of Maui in the Hawaiian island chain. Routine observations are conducted remotely, from the Waiakoa Laboratory in Pukalani.  A complete description of the PS1 system, both hardware and software, is provided by \citet{Kaiser_etal10}. The survey design and execution strategy are described in K.~C.~Chambers et al. (2011, in preparation).

The PS1 optical design \citep{Hodapp_etal04} uses a 1.8~meter diameter $f$/4.4 primary mirror, and a 0.9~m secondary.  The resulting converging beam then passes through two refractive correctors, a $48$~cm~$\times~48$~cm interference filter, and a final refractive corrector that is the dewar window. The telescope delivers an image with a diameter of 3.3 degrees, with low distortion. The PS1 imager \citep{Tonry_Onaka09} comprises a total of 60 $4800\times4800$ pixel detectors, with 10~$\mu$m pixels that subtend 0.258~arcsec, providing an instantaneous field of view of 7.1 sq. deg.  The detectors are back-illuminated CCDs manufactured by Lincoln Laboratory.  The detectors are read out using a StarGrasp CCD controller, with a readout time of 7 seconds for a full unbinned image. Initial performance assessments are presented in \citet{Onaka_etal08}. 

The PS1 observations are obtained through a set of five broadband filters, which we have designated as \gps, \rps, \ips, \zps, and \yps. Although the filter system for PS1 has much in common with that used in previous surveys, such as Sloan Digital Sky Survey \citep[SDSS;][]{York_etal00, Abazajian_etal09}, there are important differences. The \gps\ filter extends 200 \AA\ redward of $g_{\rm SDSS}$, paying the price of 5577 \AA\ sky emission for greater sensitivity and lower systematics for photometric redshifts, and the \zps\ filter is cut off at 8400 \AA, giving it a different response than the detector response defined $z_{\rm SDSS}$.  SDSS has no corresponding \yps\ filter.  Further information on the passband shapes is described in \citet{Stubbs_etal10}. Provisional response functions (including 1.3 airmasses of atmosphere) are available at the project's web site \footnote[1]{http://svn.pan-starrs.ifa.hawaii.edu/trac/ipp/wiki/ PS1\_Photometric\_System}. Photometry is in the ``natural'' PS1  system, $m=-2.5\ \textrm{log}(\textrm{flux})+m'$, with a single zeropoint adjustment $m'$ made in each band to conform to the AB magnitude scale. Zeropoints were measured from comparison with field stars in the SDSS catalog, but no color corrections were made to determine the magnitudes exactly in the SDSS system.

The PS1 Medium Deep Survey (MDS) accounts for approximately 25\% of observing time. It revisits ten fields (each equivalent to a single PS1 imager footprint) on a nearly nightly basis in \gps, \rps, \ips, \zps, and \yps\ bands \citep{Stubbs_etal10}, reaching a typical 3$\sigma$ limiting magnitude of $\sim$23.5 mag in one visit at \gps, \rps, \ips, and \zps\ bands and $\sim$21.7 mag in \yps. The MDS fields are distributed across the sky, so only a subset of fields are observed at any given time of year, and on any given night only in a subset of bands, depending on observing conditions.

The MDS images are processed through the Image Processing Pipeline \citep[IPP;][]{Magnier06}, on a computer cluster at the Maui High Performance Computer Center. The pipeline runs the images through a succession of stages, including flat-fielding (``de-trending''), a flux-conserving warping to a sky-based image plane, masking and artifact removal, and object detection and photometry. Difference images are produced from the stacked images by the \verb|photpipe| pipeline \citep{Rest_etal05} and potential transients are visually inspected by humans for possible promotion to the status of transient alerts.  The IPP-photpipe system finds hundreds of transient alerts per month, of which a subset are targeted for spectroscopic confirmation.

\subsubsection{PS1-10ky and PS1-10awh}

\begin{figure}
\centering
\includegraphics[angle=90, width=9.5cm]{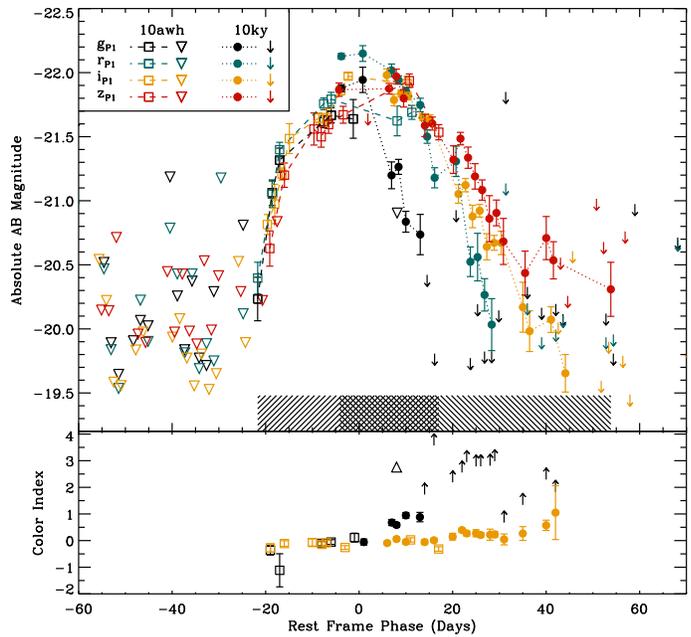}
\caption{Top Panel: Combined light curve for PS1-10awh and PS1-10ky in four filters that have been red-shifted to the central wavelengths listed in Table \ref{table:wrest}: \gps\ (black), \rps\ (blue), \ips\ (gold), and \zps\ (red). Measurements for PS1-10awh are marked as open squares connected with dashed lines; 3$\sigma$ upper limits are denoted by open triangles. For PS1-10ky, measurements are shown as filled circles connected with dotted lines, and upper limits are signified with arrows. Rest frame times and absolute magnitudes are calculated using measured redshifts of z = 0.908 and z = 0.956, respectively. To highlight the overlap of the two light curves, the time ranges during which each SN is detected in PS1 photometry are marked as hatched regions at the bottom of the plot. Bottom Panel: The time evolution of $\gps-\zps$ color in black and $\ips-\zps$ color in gold. Measurements and limits for PS1-10awh are marked as squares and triangle respectively, while measurements and limits for PS1-10ky are shown as circles and arrows.}
\label{lc_splice}
\end{figure}

PS1-10awh and PS1-10ky were discovered by the PS1 MDS at J2000 locations of R.A. = $22^{\rm h}14^{\rm m}29^{\rm s}.831$, Dec = $-00^{\circ}04^{\prime}03^{\prime\prime}.62$ and R.A. = $22^{\rm h}13^{\rm m}37^{\rm s}.851$, Dec = $+01^{\circ}14^{\prime}23^{\prime\prime}.57$, respectively (uncertainty in positions $\approx 0.05$ arcsec).  PS1-10ky was apparent immediately when observations began of the MD09 field in June 2010. It was discovered near peak brightness and had faded below PS1 MDS sensitivity by November 2010.  PS1-10awh was first detected in early October 2010 by its rising flux, and was observed almost every night by PS1 MDS until Field MD09 set in early December.  Figure \ref{comparison} shows the vicinities of the SNe before explosion and around maximum light.  Both objects were selected for spectroscopic followup at the MMT and Gemini based on the lack of visible host galaxies in the template images and were then confirmed as high-redshift supernovae.

\verb|Photpipe| measures SN photometry using forced-centroid PSF-fitting photometry on the difference images, with a PSF appropriate to each difference image, and a common centroid derived from the high SNR detections.  The measured AB magnitudes for PS1-10awh and PS1-10ky are listed in Tables \ref{table:phot_ky} and \ref{table:phot_awh} and plotted in Figure \ref{lc_app}, along with 3$\sigma$ upper limits for epochs with non-detections. We corrected for foreground extinction using \citet{Schlegel_etal98} values and the \citet{Cardelli_etal89} extinction law, but did not correct for any intrinsic extinction.

\begin{figure}
\centering
\includegraphics[angle=90, width=9.5cm]{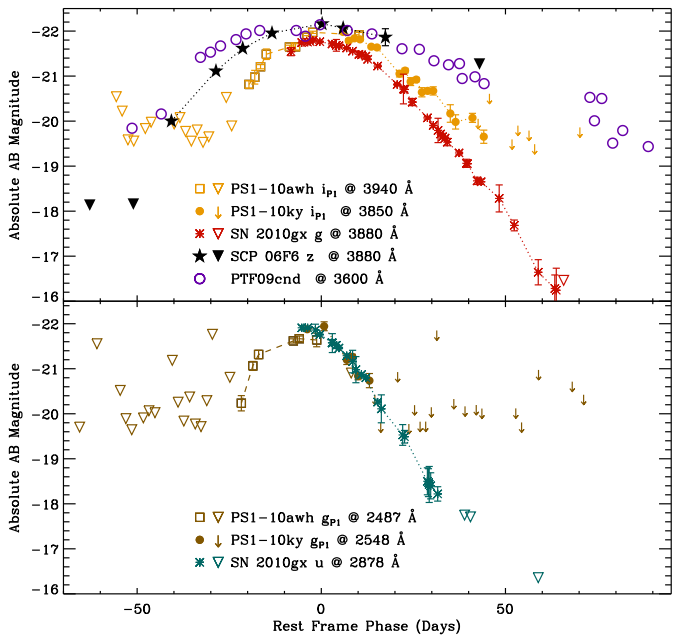}
\caption{A comparison of light curves for SCP 06F6-like ultra-luminous SNe. In the top panel, we compare \ips\ band light curves for PS1-10awh and PS1-10ky with SN 2010gx \emph{g} band measurements and SCP 06F6 \emph{z} band measurements, all of which fall at comparable rest wavelengths around 3900 \AA. Similarly in the bottom panel, we compare the \gps\ band measurements of PS1-10awh and PS1-10ky with the \emph{u} band light curve of SN 2010gx, to probe blue wavelengths around $\sim$2600 \AA\ (no bluer photometry is available for SCP 06F6).}
\label{lc_multi}
\end{figure}

To calculate absolute magnitudes, we did not carry out a full $k$-correction, but instead only corrected the measured magnitudes for cosmological expansion using the redshifts measured from \ion{Mg}{2} absorption (\S \ref{spec}):
\begin{equation}
M = m - 5\ \textrm{log}\left( {{d_{L}} \over {10\ \textrm{pc}}} \right) + 2.5\ \textrm{log}\left(1 + z\right)
\end{equation}
where $m$ is the apparent magnitude, $d_{L}$ is the luminosity distance, and $M$ is the corresponding absolute magnitude. We did not shift the central wavelengths at which flux densities were measured (see Table \ref{table:wrest} for central wavelengths in the rest frame, which are weighted by the system response). The similar redshifts of PS1-10ky (z = 0.956) and PS1-10awh (z = 0.908) ensure that the rest wavelengths and passbands of the photometry will be comparable. 

The light curves show similar peak luminosities, widths, and color evolution. We find that we can splice them together to produce a full light curve if we set the peak date to 2010 July 20 for PS1-10ky and to 2010 November 20 for PS1-10awh (Figure \ref{lc_splice}). For both objects, we observe the peak of the light curve, implying that PS1-10awh fully represents the light curve rise and PS1-10ky samples the light curve decline. Uncertainties in the splicing (or in our implicit assumption that these two sources are very similar) will only affect \emph{comparisons} of the rise and the decline. We have overlapping light curve coverage for $\sim$18 rest-frame days around peak (see hatched region in Figure \ref{lc_splice}). The splicing of the light curves is well-constrained by a few late-time measurements of PS1-10awh: noteably the \zps\ measurement from 2010 December 10 and the \gps\ upper limit from 2010 December 5, which both show a declining light curve (Figure \ref{lc_app}). Note the well-matched rapid fall off in the \gps\ band for both sources (Figure \ref{lc_splice}), implying that they have a similar color evolution.

\begin{deluxetable}{lcccc}
\tablewidth{0 pt}
\tablecaption{ \label{table:wrest}
 Central Rest Wavelengths of Optical Passbands (\AA) }
\tablehead{Filter\tablenotemark{a} & PS1-10awh & PS1-10ky & SN 2010gx & SCP 06F6 }
\startdata
$u$ &  ---    & ---      & 2878 & --- \\
$g$ & 2550 & 2488 & 3878 & --- \\
$r$  & 3267 & 3188 & 5065 & --- \\
$i$  & 3944 & 3848 & 6199 & 3540 \\
$z$ & 4536 & 4426 & 7426 & 3883 \\
$y$ & 5097 & 4974 & --- & --- \\
\enddata
\tablenotetext{a}{These are PS1 filters \citep{Stubbs_etal10} for PS1-10ky and PS1-10awh, SDSS filters for SN 2010gx, and \emph{Hubble Space Telescope}/ACS filters \emph{F775W} and \emph{F850LP} for SCP 06F6.}
\end{deluxetable} 

In Figure \ref{lc_multi}, we compare the combined light curve from PS1-10ky and PS1-10awh with the ultra-luminous transient SCP 06F6 (\citealt{Barbary_etal09, Quimby_etal11}; z = 1.19) and SN 2010gx (\citealt{Pastorello_etal10}; z = 0.23). The combination of our two PS1 SNe show a light curve that falls off much more slowly at the redder ($\sim$3900 \AA; gold points) bands, as compared with the bluer ($\sim$2500 \AA; black points) measurements. SN 2010gx was not well observed on the rise, although the one meaningful upper limit (unfiltered; \citealt{Pastorello_etal10}) is suggestive of a faster rise to maximum than the other three. Its decline shows the bluer filter declining much faster than the redder, in good accord with PS1-10ky and PS1-10awh. The sources' peak luminosities agree to within $\sim$25\% and the widths of the light curves are similar. From the commonalities in their light curves, along with the similarities in their spectra (\S \ref{spec}), we conclude that these four sources all belong to the  common class of SCP 06F6-like ultra-luminous SNe pointed out by \citet{Quimby_etal11}.

\subsection{Spectroscopy} \label{spec}

\begin{deluxetable*}{lccccccc}
\tablewidth{0 pt}
\tablecaption{ \label{table:spec}
 Spectroscopic Observations}
\tablehead{Date & Phase\tablenotemark{a} & Facility & Grating/ & $\lambda$ Range\tablenotemark{b} & Res\tablenotemark{c} & Airmass & Slit P.A. \\
 & (Days) & & Central $\lambda$ (\AA) & (\AA) & (\AA) & & (deg) }
\startdata
\underline{PS1-10ky:} & & & & & & & \\
2010 Jul 17 & $-$2 & MMT/Blue Channel\tablenotemark{d} & 300/6006 & 3393--8623 & 5.5 & 1.2 & 326 \\
2010 Jul 21 & 1  & Gemini-N/GMOS\tablenotemark{e} & B600/4800 & 3390--6220 & 4.7 & 1.1 & 180 \\
2010 Aug 18  &15 & Gemini-S/GMOS & B600/5000 & 3600--6432 & 4.7 & 1.3 & 180 \\
2010 Sep 09 & 26 & Gemini-S/GMOS & R400/8000 & 5887--10161 & 8.0 & 1.4 & 180 \\
& & & & & & & \\
\underline{PS1-10awh:} & & & & & & & \\
2010 Oct 12 & $-$21 & Gemini-N/GMOS & R400/7500 & 5441--9655 & 5.0 & 1.3 & 210 \\
2010 Nov 27 & 4 & MMT/Hectospec\tablenotemark{f} & 270/6500 & 3700--9150 & 4.9 & 1.2 & Fiber \\
2010 Dec 09 & 10 & MMT/Blue Channel & 300/5768 & 3230--8325 & 5.5 &1.5 & 40 \\
\enddata
\tablenotetext{a}{In the rest frame.}
\tablenotetext{b}{In the observer frame.}
\tablenotetext{c}{Approximate spectral resolution, measured from widths of night sky lines.}
\tablenotetext{d}{\citet{Schmidt_etal89}. $^{e}$\citet{Hook_etal04}. $^{f}$\citet{Fabricant_etal05}.}
\end{deluxetable*} 

\begin{figure}
\centering
\includegraphics[angle=90, width=9.5cm]{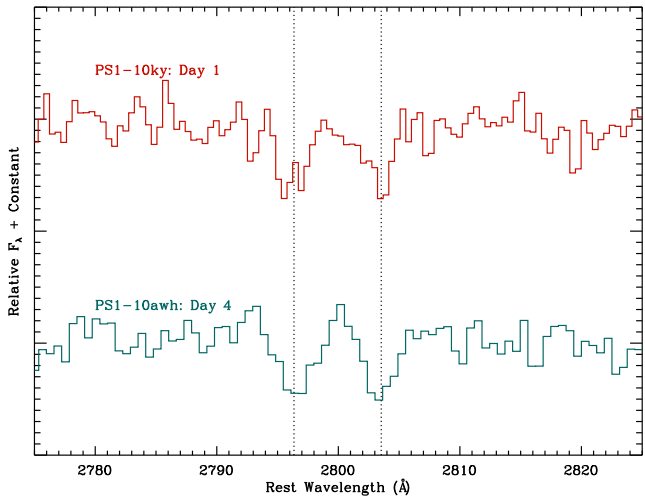}
\caption{Detail of two spectra with wavelength coverage of the \ion{Mg}{2} $\lambda\lambda$2796,2803 doublet, redshifted to z = 0.9084 for PS1-10awh and z = 0.9558 for PS1-10ky. The rest wavelengths of the \ion{Mg}{2} doublet are shown as vertical dotted lines.}
\label{mgii}
\end{figure}

We obtained four optical spectra of PS1-10ky and three spectra of PS1-10awh using MMT (Hectospec and Blue Channel Spectrographs) and Gemini (GMOS; Table \ref{table:spec}). Basic spectroscopic processing and extraction were accomplished using standard routines in IRAF, or using the OIR Telescope Data Center pipeline in the case of MMT/Hectospec data \citep{Mink_etal07}.  We then used custom IDL routines to apply flux calibrations and remove telluric absorption based on observations of spectrophotometric standard stars. Narrow \ion{Mg}{2} $\lambda\lambda$2796,2803 absorption, arising from the interstellar medium of the SN host galaxy, is seen in all spectra with wavelength coverage of the doublet (Figure \ref{mgii}). From this absorption, we measure a redshift of z = 0.9084 for PS1-10awh and z = 0.9558 for PS1-10ky. These correspond to luminosity distances of d$_{L}$ = 5890 Mpc and  d$_{L}$ = 6270 Mpc for a cosmology with H$_{0}$ = 71 km s$^{-1}$ Mpc$^{-1}$, $\Omega_{M}$ = 0.27, and $\Omega_{\Lambda}$ = 0.73.

\begin{figure}
\centering
\includegraphics[angle=90, width=9.5cm]{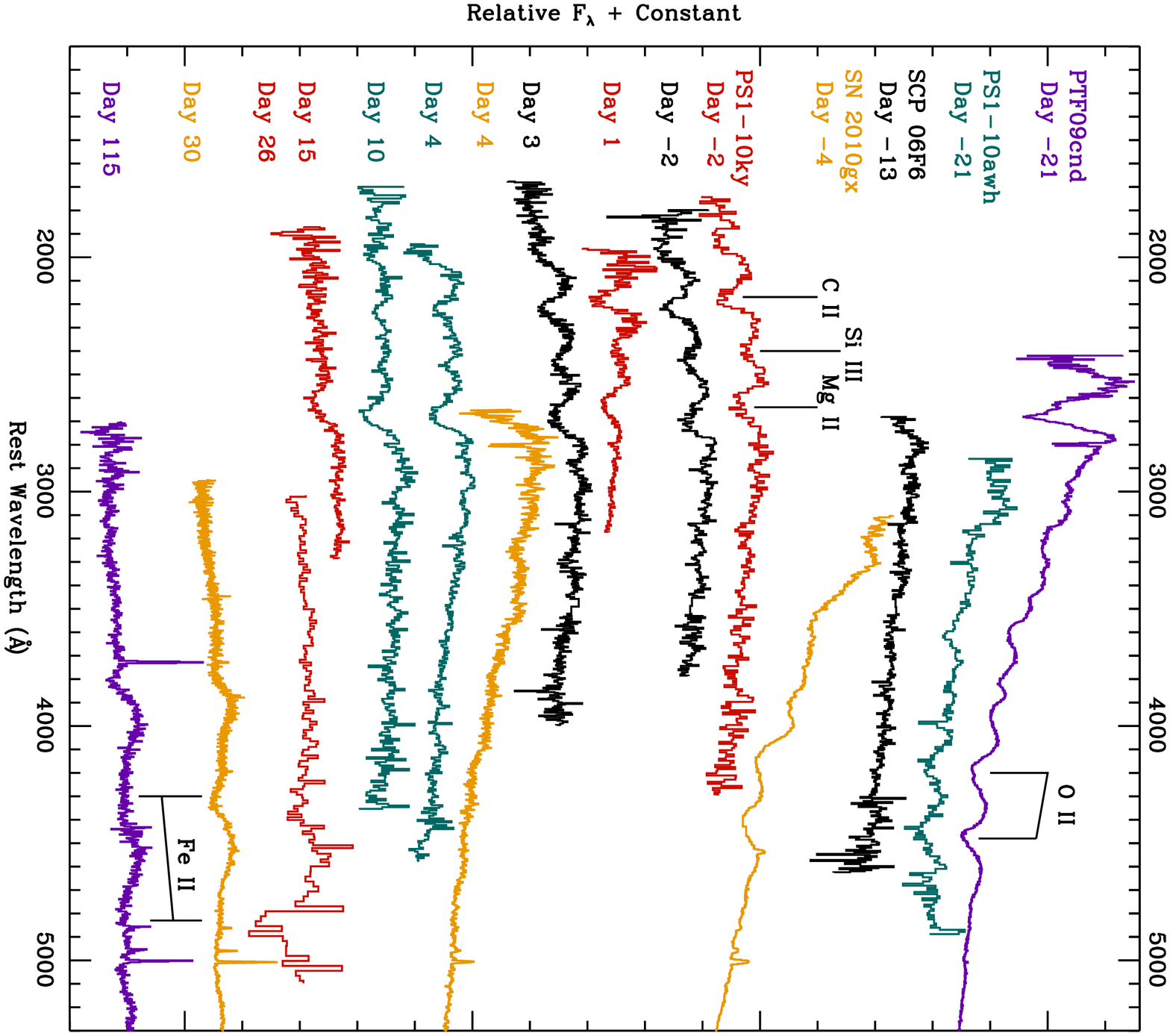}
\caption{Spectra of PS1-10awh in blue, PS1-10ky in red, SN 2010gx \citep[from][]{Pastorello_etal10} in gold, and SCP 06F6 \citep[from][]{Barbary_etal09} in black. The time of each spectrum, relative to light curve peak in the rest frame, is marked on the left. Also marked are strong spectral features, tenatively identified as \ion{C}{2}, \ion{O}{2}, \ion{Si}{3}, \ion{Mg}{2}, and \ion{Fe}{2} as discussed in the text.}
\label{spec_multi}
\end{figure}

The spectra are shown in Figure \ref{spec_multi}, along with the SCP 06F6 spectra from \citet{Barbary_etal09} and spectra of SN 2010gx from \citet{Pastorello_etal10}. As with the photometry, we corrected them for foreground extinction \citep{Schlegel_etal98, Cardelli_etal89}. In many cases, clouds were present during observations so the absolute flux scaling of the spectra can not be trusted, but the general shape of the continuum is reliable. 

The three broad absorption features between 2000 and 2900 \AA\ were identified by \citet{Quimby_etal11} as \ion{C}{2} ($\sim$2330 \AA), \ion{Si}{3} (2543 \AA), and \ion{Mg}{2} (2800 \AA). The broad ``W''-shaped features around 4300 \AA, identified by \citet{Quimby_etal11} as \ion{O}{2}, can be seen in the GMOS spectrum of PS1-10awh observed on Day $-21$.

Note that PS1-10ky and PS1-10awh show spectral features that become weaker as time progresses; our later spectra on PS1-10ky appear almost featureless. In SN 2010gx, the 4300 \AA\ ``W'' feature has disappeared and the spectrum is practically featureless by Day 4, but then subsequently it shows strong SN Ic-like features of \ion{Fe}{2}, \ion{Ca}{2}, and \ion{Mg}{2}. We are not very sensitive to similar features in PS1-10ky and PS1-10awh because of our blue rest-frame wavelength coverage (Figure \ref{spec_multi}). Our last spectrum, observed with relatively red wavelength coverage, was observed 26 days after peak and is rather noisy, but may show low signal-to-noise ratio depressions around rest wavelengths of 4350 \AA\ and 4900 \AA\ which might be consistent with the \ion{Fe}{2} and \ion{Mg}{2} features in SN 2010gx. 

It is remarkable that no lines of H or He have ever been detected from an SCP 06F6-like object, although we note that  none of the spectra in Figure \ref{spec_multi} cover H$\alpha$ and the few that provide coverage of H$\beta$ are very noisy. Similarly with He, the strongest lines are redward of 5000 \AA\ and outside our spectroscopic coverage. However, the spectra obtained on low-redshift SCP 06F6-like objects by \citet{Quimby_etal11} and \citet{Pastorello_etal10} do cover this red portion of the spectrum, and no trace of H or He has been detected in these studies. In the future, deep infrared spectroscopy will be critical for placing strong limits on the presence of H or He.

\subsection{GALEX Photometry}\label{galex}

MD09 was observed with \emph{GALEX} in August--September 2010. Both PS1-10ky and PS1-10awh---and their hosts---were non-detections in the NUV band. PS1-10ky was observed by \emph{GALEX} near optical peak brightness, but PS1-10awh was observed before its rise. NUV 3$\sigma$ upper limits are listed in Tables \ref{table:phot_ky} and \ref{table:phot_awh}. The upper limits of $\gtrsim$ 23.5 mag (AB system) correspond to a luminosity limit of $\lesssim 1.0 \times 10^{41}$ erg s$^{-1}$ \AA$^{-1}$ at a central rest wavelength of 1161 \AA\ for PS1-10ky.

\subsection{EVLA Radio Continuum Measurements}

We observed both ultra-luminous SNe with the EVLA as part of our NRAO Key Science Project ``Exotic Explosions, Eruptions, and Disruptions: A New Transient Phase-Space''. We observed at 4.9 GHz with 256 MHz of bandwidth; time on source was 37 minutes for both sources. Gains were calibrated using J2212+0152, and the absolute flux density scale was calibrated using 3C48. PS1-10awh was observed on 2010 Dec 12, approximately three weeks after optical peak (observer frame), and yielded a non-detection of $-22\pm15\ \mu$Jy beam$^{-1}$. PS1-10ky was observed on 2011 Feb 12, almost seven months after peak; we measured a non-detection of $-2\pm17\ \mu$Jy beam$^{-1}$ at its location. 

The gamma-ray burst (GRB) 030329 would have had a peak 4.9 GHz flux density of $\sim$200 $\mu$Jy at the distance of our ultra-luminous SNe, and would have remained a $>3\sigma$ source for $\sim$100 days \citep{vanderHorst_etal05, Frail_etal05}. On the other hand, a weak GRB like 980425 and its associated SN 1998bw would be significantly below our detection limit, with a 4.8 GHz peak flux density of 2 $\mu$Jy at a distance of 6000 Mpc \citep{Kulkarni_etal98}. It is unlikely, although not impossible given the spread in GRB radio luminosities \citep{Soderberg07}, that either PS1-10awh or PS1-10ky hosted a GRB.

\section{Host Galaxies} \label{host}

We stacked the pre-explosion images of PS1-10awh (prior to 2010 Sept 21) to derive upper limits on the host galaxy's luminosity.  We find 3$\sigma$ upper limits on apparent magnitude of  $\gps > 26.6$, $\rps > 26.6$, $\ips > 26.7$, $\zps > 26.3$, and $\yps > 24.2$ mag. Our $\zps$-band limit roughly corresponds to M$_B > -17.0$ mag or $<$ 0.02 L$^{\star}$ (assuming M$_{B}^{\star} = -21.3$ mag at a redshift of 0.9; \citealt{Faber_etal07}). Similarly, we stacked the 2009 data for PS1-10ky and found apparent magnitude limits of $\gps > 26.1$, $\rps > 26.0$, $\ips > 25.9$, $\zps > 25.3$, and $\yps > 23.3$ mag. The $\zps$-band limit translates to M$_B > -18.0$ mag or $<$ 0.05 L$^{\star}$.


The $\gps$ band is centered at $\sim$2500 \AA, in the NUV spectral range commonly used as a tracer of the photospheres of young massive stars, and thereby star formation rate (SFR). Using the calibration of \citet{Kennicutt98} and assuming no dust exctinction, our limits on the $\gps$-band host galaxy luminosities translate to SFR $<$ 0.4 M$_{\odot}$ yr$^{-1}$ for PS1-10ky and $<$ 0.3 M$_{\odot}$ yr$^{-1}$ for PS1-10awh. We also stacked the three spectra for PS1-10awh and detected [\ion{O}{2}]$\lambda$3727 emission at a luminosity of $4.5\pm1.0 \times 10^{40}$ erg s$^{-1}$. This translates to a SFR of $0.4 \pm 0.2$ M$_{\odot}$ yr$^{-1}$ \citep{Kewley_etal04}, consistent with the NUV limits given the significant uncertainties and our poor understanding of dust extinction in these systems. We did not stack the PS1-10ky spectra, as the redshift of this source places the [\ion{O}{2}] emission line on a telluric feature.

These constraints are consistent with the host galaxies of other ultra-luminous SNe \citep{Neill_etal11}, which typically have low masses (less than the Large Magellanic Cloud), relatively low SFRs (70\% of the Neill et al. sample has SFR $< 1$ M$_{\odot}$ yr$^{-1}$), and high specific SFRs. This preference for low-mass hosts is reminiscent of GRBs, which also display a bias towards lower-mass hosts with high specific SFR \citep[e.g.,][]{LeFloch_etal03, Christensen_etal04}. In the context of GRBs, this bias is often attributed to a collapsar origin \citep{MacFadyen_Woosley99}. The mass--metallicity relationship for galaxies \citep{Tremonti_etal04} implies that lower-mass galaxies will be more metal poor, and metal-poor SN progenitors will undergo less wind-driven mass loss and maintain more angular momentum. However, a recent alternative hypothesis has been offered in which, because low-mass galaxies have higher specific SFRs, the observed bias of GRBs towards low-mass hosts is in fact simply a demonstration that they are associated with young stellar populations \citep{Kocevski_West11, Mannucci_etal11}. 

\section{Color Evolution}\label{temp}

\begin{deluxetable*}{lccccc}
\tablewidth{0 pt}
\tablecaption{ \label{table:spec2}
 Spectroscopic Measurements}
\tablehead{Date & Phase\tablenotemark{a} & Temperature & Velocity\tablenotemark{b} & FWHM\tablenotemark{c}\\
 & (Days) & (K) & (10$^{3}$ km s$^{-1}$) & (10$^{3}$ km s$^{-1}$)}
\startdata
\underline{PS1-10ky:} & & \\
2010 Jul 17 & $-$2 & 13,400$\pm$2700 & 19 & 12 \\
2010 Jul 21 & 1  & 18,400$\pm$9000 & 20 & 13 \\
2010 Aug 18  &15 & 9600$\pm$4800 & 19 & 14 \\
2010 Sep 09 & 26 & 7400$\pm$1100 & --- & --- \\
 & & \\
\underline{PS1-10awh:} & & \\
2010 Oct 12 & $-$21 &18,100$\pm$3600 & 15\tablenotemark{d} & 11\tablenotemark{d}\\
2010 Nov 27 & 4 & 19,700$\pm$3900 & 11 & 9 \\
2010 Dec 09 & 10 & 10,200$\pm$2000 & 11 & 9 \\
 & & \\
\underline{SCP 06F6:\tablenotemark{e}} & & \\
2006 Apr 22 & $-$13 & 19,800$\pm$4000 & --- & ---\\
2006 May 18 &$-$2 & 13,000$\pm$2600 & 12 & 12 \\
2006 May 28 & 3 & 13,600$\pm$1400 & 13 & 12 \\
 & & \\
\underline{SN 2010gx:\tablenotemark{f}} & & \\
2010 Mar 21 & $-$5 & 16,300$\pm$1600 & 19\tablenotemark{d} & 12\tablenotemark{d}\\
2010 Mar 22 &$-$4 & 15,200$\pm$800 & 19\tablenotemark{d} & 9\tablenotemark{d} \\
2010 Apr 01 & 4 & 15,000$\pm$2300 & --- & --- \\
2010 Apr 09 & 10 & 12,900$\pm$1900 & 19\tablenotemark{g} & 12\tablenotemark{g} \\
2010 Apr 22 &21 & 10,400$\pm$1600 & 17\tablenotemark{g} &16\tablenotemark{g} \\
2010 May 02 & 30 & 6900$\pm$1000 & 15\tablenotemark{g} & 9\tablenotemark{g} \\
2010 Jun 05 & 57 & 5100$\pm$500 & --- & --- \\
\enddata
\tablenotetext{a}{In the rest frame.}
\tablenotetext{b}{Velocity corresponding to line center for the three features in the blue (rest wavelengths of $\sim$2330, 2540, and 2800 \AA) unless otherwise noted. Typical errors of $\sim$1000 km s$^{-1}$.}
\tablenotetext{c}{Average FWHM for the the three features in the blue, unless otherwise noted. Typical errors of $\sim$3000 km s$^{-1}$}
\tablenotetext{d}{This is a red spectrum, so velocities were measured from the \ion{O}{2} ``W'' feature at $\sim$4300 \AA.}
\tablenotetext{e}{Spectra from \citet{Barbary_etal09}.}
\tablenotetext{f}{Spectra from \citet{Pastorello_etal10}.}
\tablenotetext{g}{Measured using the broad \ion{Fe}{2}+\ion{Mg}{2} feature around 4300 \AA.}
\end{deluxetable*}

\begin{figure}
\centering
\includegraphics[angle=90, width=9.5cm]{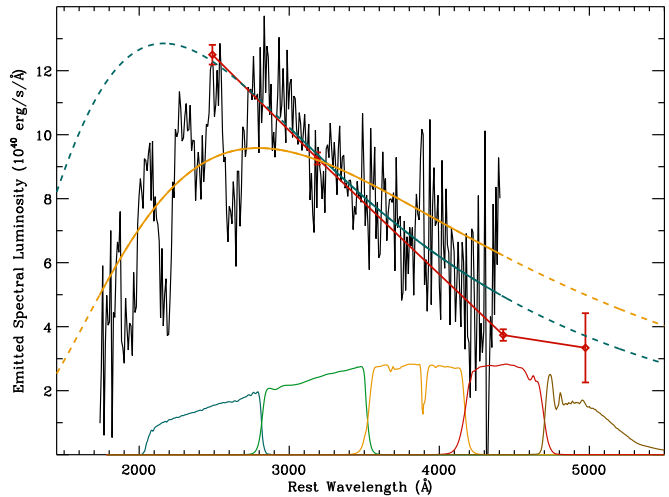}
\caption{Spectrum of PS1-10ky observed at a phase of $-2$ days. Our PS1 photometry is overlain as red points connected with lines; the system response functions of the \gps, \rps, \ips, \zps, and \yps\ filters are also shown at the bottom of the plot for reference. A blackbody fit to the full spectrum (T = 10,000 K) is shown as a solid gold line, and its extension to other wavelengths is represented by the gold dashed line. Similarly, the blackbody fit to the spectrum redward of 3000 \AA\ (T = 13,000 K) is shown as a blue line.}
\label{bb}
\end{figure}

\begin{figure}[htp]
\centering
\includegraphics[angle=90, width=9.5cm]{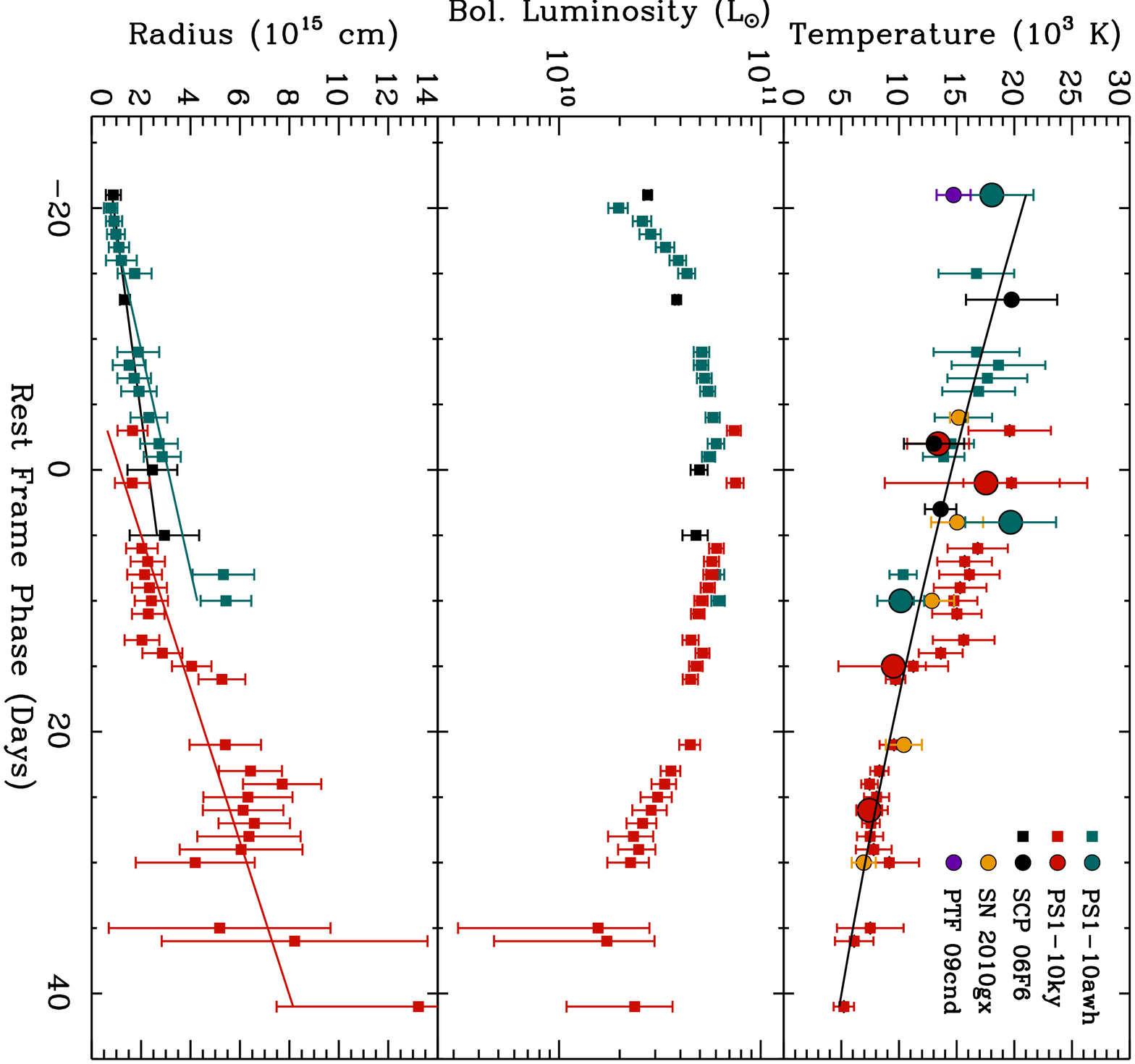}
\caption{Top Panel: Temperature as a function of rest-frame time. Squares denote measurements from photometry, while circles are spectroscopic measurements. The quadratic fit to these data is marked as a black solid line. Middle Panel: Bolometric luminosity as a function of time. These are determined by summing up flux in the measured photometric bands and extrapolating a blackbody tail to the red; emission blueward of 2100 \AA\ for PS1-10ky/PS1-10awh, 2600 \AA\ for SN 2010gx, and 3150 \AA\ for SCP 06F6 is not accounted for, implying that these are lower limits. Bottom Panel: Radius as a function of time, assuming the bolometric luminosities from the middle panel and the temperatures from the top panel. Linear fits are marked by solid lines. In all three panels, measurements for PS1-10awh are in blue, PS1-10ky in red, SCP 06F6 in black, and SN 2010gx in gold.}
\label{t_l_r}
\end{figure}

In order to understand the basic physical parameters of these ultra-luminous SNe, we need to constrain their temperature evolution. The SEDs of SCP 06F6-like objects can be reasonably fit as blackbodies at redder wavelengths, but appear to suffer severe line blanketing blueward of 3000 \AA. This was pointed out for SN 2010gx by \citet{Pastorello_etal10} and is apparent in Figure \ref{bb}, where we show a spectrum of PS1-10ky with a blackbody fit to the full spectrum (T = 10,400 K) and a fit to the spectrum redward of 3000 \AA\ (T = 13,400 K). Clearly, no good blackbody fit is achievable when the bluest wavelengths are included, and the luminosity at $\lesssim$ 2500 \AA\ is significantly damped from what is expected for the blackbody fit to the redder wavelengths. All of the spectra in Figure \ref{spec_multi} turn over blueward of 3000 \AA, so we fit blackbodies to our data redward of 3000 \AA. 

We carried out least-squares fits of Planck functions to the spectra shown in Figure \ref{spec_multi}; the best-fit temperatures and errors are listed in Table \ref{table:spec2}. We also fit color temperatures to the photometric data points for PS1-10awh, PS1-10ky, and SN 2010gx by interpolating the measured fluxes in time with a cubic spline and then least-squares fitting a blackbody spectrum on each rest-frame day which is constrained by measurements at three or more bands with central wavelengths redward of 3000 \AA. An interpolated flux on a given day is considered ``constrained" if there is a $>$3$\sigma$ measurement both preceding it and following it in time. At times when there are only two bands with constrained flux measurements redward of 3000 \AA, we fit the colors by calculating a grid of predicted colors for various blackbody temperatures (grid resolution of 250 K) and identifying the closest blackbody temperature to the measured color.  Errors in color are directly translated to errors in temperature by using this grid in a similar fashion. Temperature fits are shown in the top panel of Figure \ref{t_l_r}. Jumps in temperature or in the size of the error bars (e.g., around Day $-7$ for PS1-10awh) are caused by transitions from least-squares fitting to color fitting; single color fits have much larger error bars and in some temperature regimes, they show systematic bias. This is likely due to features in the spectra which can cause significant deviations from a blackbody spectrum over several hundred angstroms (Figure \ref{spec_multi}). 

The spectroscopic and photometric temperature measurements agree rather well, and the color temperatures of all four ultra-luminous SNe are consistent with one another. The temperature evolution is modeled with a quadratic fit to the combined measurements for all four SNe (solid black line in Figure \ref{t_l_r}) showing a maximum temperature of $\sim$24,000 K at early times (with significant uncertainty) and a minimum temperature of $\sim$5000 K at late times. 

From Figure \ref{bb}, we see that even in the absence of line blanketing, we would not expect GALEX detections with $< 23.5$ mag (corresponding to $\lesssim 1.0 \times 10^{41}$ erg s$^{-1}$ \AA$^{-1}$ at 1161 \AA). The first GALEX observation of PS1-10ky occurred at $\sim$6 days after maximum, when the temperature was cooler and the ultraviolet luminosity would have been even fainter than that predicted in Figure \ref{bb}.

We note that the \ion{C}{2} and \ion{Si}{3} absorption have largely disappeared by Day 10, whereas the \ion{Mg}{2} absorption is still strong (Figure \ref{spec_multi}). This might be explained by a cooling photosphere, as \ion{Mg}{2} has a lower ionization potential (7.6 eV) than \ion{C}{2} (11.3 eV) or  \ion{Si}{3} (16.4 eV). Similarly, the \ion{O}{2} lines have become weaker by Day 4, consistent with its rather high ionization potential of 13.6 eV. Also, the source of the line blanketing blueward of 3000 \AA\ remains an interesting mystery, as line blanketing is usually caused by Fe-peak elements, but there are no iron lines apparent in the spectra during or preceding light curve maximum.

\section{Bolometric Luminosity}\label{lum}

We totaled up the interpolated flux for PS1-10ky before Day $-$4 and PS-10awh after Day $-$4, and found a radiated energy of $(6.2 \pm 0.7) \times 10^{50}$ erg. This was calculated using trapezoidal interpolation across all observed bands, and linearly extrapolating the integration out to the edges of the observed bands (most importantly, we impose a blue cut-off of 4050 \AA\ to the \gps\ band in the observer frame). We can perform a similar calculation for the multi-band photometry of SN 2010gx, which covers times after Day $-12$ and a rest-frame wavelength range of 2600--7300 \AA\ when $u$-band measurements are included (SCP 06F6 is not detected in a sufficient number of bands to justify such an analysis). For SN 2010gx we measure a total radiated energy of $(3.7 \pm 0.2) \times 10^{50}$ erg. These are strong lower limits on the total radiated energy because they only represent the measured flux. Not all bands are constrained at all dates, and even when fluxes are constrained at all bands, we are only measuring the SED over a limited wavelength range ($\sim$2100--5300 \AA\ in the rest frame for PS1-10ky and PS1-10awh). Conveniently, the SEDs of SCP 06F6-like objects peak in the ultraviolet, around $\sim$2500 \AA\ (e.g., Figures \ref{spec_multi} and \ref{bb}), so despite our limited rest-frame wavelength coverage, we may still be detecting the majority of the radiated energy. 

Next, we attempt to more realistically model the bolometric luminosity by accounting for the flux emitted redward of the reddest band measured. The behavior of the SED blueward of the $\gps$ band is poorly understood and clearly diverges from a blackbody (Figure \ref{bb}), but at redder wavelengths the spectrum can be roughly described as a Planck law. On any given date, we assume a temperature from the polynomial fit to our data (black line in top panel of Figure \ref{t_l_r}; the actual data imply error bars on temperature that are so large at some times, they preclude a meaningful measurement of bolometric luminosity). We then splice the red tail of a blackbody curve of this temperature onto the reddest constrained photometric point, and sum up the measured photometric points (as described above) with this blackbody tail. Bolometric light curves estimated using this technique are plotted in the middle panel of Figure \ref{t_l_r}. The bolometric light curves of PS1-10awh and PS1-10ky match up rather well, with maximum luminosities of $(2.3 \pm 0.3) \times 10^{44}$ erg s$^{-1}$ and $(3.0 \pm 0.4) \times 10^{44}$ erg s$^{-1}$ respectively, translating to a bolometric magnitude of $-22.5$ mag. For SN 2010gx and SCP 06F6, the light curves peak at $(1.8 \pm 0.1) \times 10^{44}$ erg s$^{-1}$ and $(2.0 \pm 0.9) \times 10^{44}$ erg s$^{-1}$ respectively, but bear in mind that these are lower limits. The bolometric light curves are also less symmetrically shaped than the single-band light curves, displaying a slower decline than rise. This is due to the cool temperatures at late times, which imply a larger correction from the red tail of the blackbody curve. 

Using this technique, we find an integrated radiated energy of $(8.7 \pm 1.0) \times 10^{50}$ erg for the combined light curve of PS1-10ky and PS1-10awh. This again is likely to be a lower limit, as we are not accounting for any flux blueward of 2100 \AA\, and after Day 16, when the $\gps$-band flux fades below detectability, we are not accounting for flux blueward of $\sim$2850 \AA. Similarly, for SN 2010gx we measure $(4.3 \pm 0.2) \times 10^{50}$ erg redward of 2600 \AA\ and after day $-$5. Finally, we measure a radiated energy of $(5.9 \pm 1.6) \times 10^{50}$ for SCP 06F6; in this case we only measure flux redward of 3150 \AA\ in the rest frame, and between Days $-$21 and 17 (before Day $-$21, there are no constraints on the temperature evolution). This technique for estimating bolometric luminosity provides lower limits on all four sources, but these lower limits approach the true bolometric luminosity most closely for PS1-10ky and PS1-10awh, where we have good blue sensitivity out to 2100 \AA\ in the rest frame. For SCP 06F6, assuming a similar color evolution as PS1-10ky/PS1-10awh in the ultraviolet, we are likely underestimating its bolometric luminosity by a factor of $\sim$2.

If we assume that these ultra-luminous SNe can be described as blackbodies of the temperatures measured in \S \ref{temp} across their entire spectra (or alternatively, that the line-blanketed blue flux re-emerges at redder wavelengths, conserving the blackbody luminosity), the estimate of the total radiated energy for PS1-10ky/PS1-10awh increases to $(2.6^{+6.2}_{-1.9}) \times 10^{51}$ erg; note the extremely large error bars due to very uncertain temperature determinations on some days. More solid temperature determinations can be made for SN 2010gx, but only after Day $-$11. We measure a total radiated energy of $(1.0 \pm 0.1) \times 10^{51}$ erg over this limited time range for this source. If instead we use our quadratic fit to the temperature evolution, the integrated radiated energy of PS1-10ky/PS1-10awh is $(1.4 \pm 0.6) \times 10^{51}$ erg, for SN 2010gx it is $(1.1 \pm 0.5) \times 10^{51}$ erg, and the total radiated energy of SCP 06F6 is $(1.6 \pm 0.8) \times 10^{51}$ erg. These can essentially be viewed as upper limits on the radiated energy over the time ranges constrained by observations.

\section{Expansion Velocity}\label{vel}

\begin{figure}
\centering
\includegraphics[angle=90, width=9.5cm]{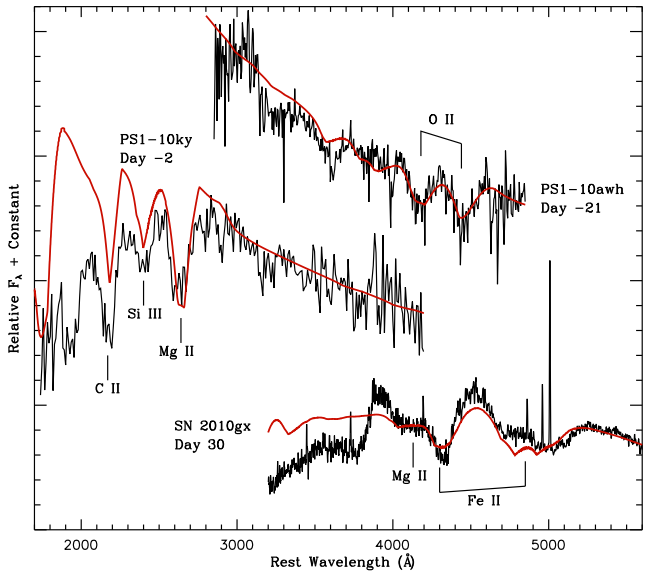}
\caption{SYNOW fits to three spectra are marked as red lines overlying the spectra. The top spectrum is of PS1-10awh on Day $-$21, and was fit with \ion{O}{2}. The middle spectrum is of PS1-10ky on Day $-$2 and shows the characteristic blue triplet of lines, fit with \ion{C}{2}, \ion{Si}{3}, and \ion{Mg}{2}. The bottom spectrum is SN 2010gx on Day 30, and is fit with \ion{Fe}{2} and \ion{Mg}{2}.}
\label{spec_synow}
\end{figure}

\begin{figure}[htp]
\centering
\includegraphics[angle=90, width=9.5cm]{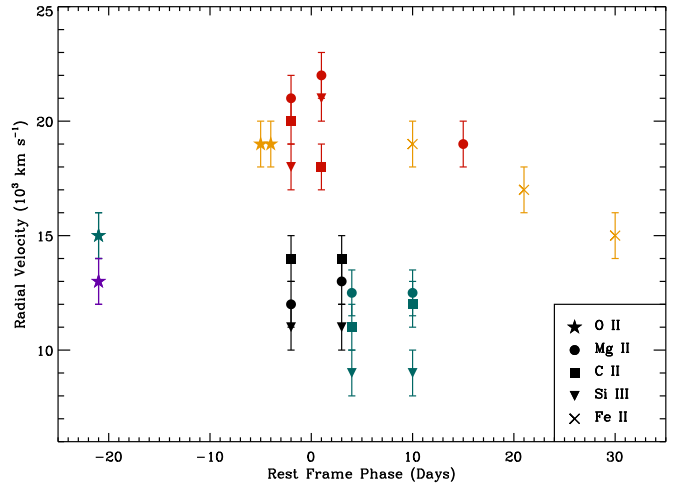}
\caption{Radial velocities measured from absorption lines as a function of rest-frame time relative to light-curve maximum. Measurements for PS1-10awh are in blue, PS1-10ky in red, SCP 06F6 in black, and SN 2010gx in gold. Different absorption lines are denoted by a variety of symbols, as shown in the key at lower right.}
\label{radvel}
\end{figure}

We can constrain the expansion velocity of these SNe by measuring the blueshift of the absorption lines, assuming that their rest frame coincides with the narrow \ion{Mg}{2} absorption seen in their spectra (presumably from the host galaxy), and that the lines are identified correctly in \citet{Quimby_etal11}. For spectra with coverage in the blue, we modeled \ion{C}{2}, \ion{Si}{2}, and \ion{Mg}{2} in \verb|SYNOW| \citep{Jeffery_Branch90}, with lines formed at an excitation temperature of 10,000 K, a radial power-law distribution of line optical depth proportional to $r^{-7}$, and a maximum ejecta velocity of 40,000 km s$^{-1}$. A typical fit is shown as the middle spectrum in Figure \ref{spec_synow}; clearly there is additional line blanketing that is not being characterized by the simple \verb|SYNOW| fit, but the absorption features are roughly reproduced. The central wavelength of each absorption feature is primarily determined by the \verb|SYNOW| parameter \verb|vmine|, which signifies the lowest velocity at which a given ion is present in the ejecta. In all cases, values of \verb|vmine| for \ion{C}{2}, \ion{Si}{2}, and \ion{Mg}{2} agree within 4000 km s$^{-1}$ (Figure \ref{radvel}, and we set the photospheric velocity to the lowest value of \verb|vmine| found between these three ions. For the expansion velocities quoted in Table \ref{table:spec2}, we take the average of the three values of \verb|vmine|. 
We find expansion velocities measured at absorption minima of $\sim$19,000 km s$^{-1}$ for PS1-10ky and $\sim$12,000 km s$^{-1}$ for PS1-10awh and SCP 06F6 (Table \ref{table:spec2}). A glance at Figure \ref{spec_multi} confirms that PS1-10ky is indeed expanding faster than the other two SNe---the three absorption features between 2000--3000 \AA\ are noticeably bluer for this source. 

For the Day $-$21 spectrum of PS1-10awh with redder wavelength coverage, we similarly fit the \ion{O}{2} ``W''-shaped feature in \verb|SYNOW|, making the same assumptions as above except using an excitation temperature of 13,000 K (Figure \ref{spec_synow}). We find an expansion velocity of $\sim$15,000 km s$^{-1}$. We also fit the expansion velocity of SN 2010gx using this feature in the Day $-$5 and Day $-$4 spectra, and found a velocity of $\sim$19,000 km s$^{-1}$ in both cases. Later in the evolution of SN 2010gx, the \ion{O}{2} ``W''-shaped feature disappears, and we fit the velocity in \verb|SYNOW| using the broad absorption feature around 4300 \AA\ on Days 10, 21, and 30, describing it with a blend of \ion{Fe}{2} and \ion{Mg}{2} and assuming basic parameters as described above (bottom spectrum in Figure \ref{spec_synow}). The spectrum of SCP 06F6 from 2006 April 22 and the spectrum of PS1-10ky from 2010 Sep 9 could not be evaluated, as there are no clear spectral features present (we note that there are no apparent \ion{O}{2} features in the SCP 06F6 spectrum from Day $-$13, while they are clear in the first spectrum of PS1-10awh from Day $-$21 and the SN 2010gx spectra from Day $-$5 and Day $-$4).

As a simple diagnostic of line shape, we also fit the FWHM of each line and take the average if there are multiple features measured in a spectrum; these are also listed in Table \ref{table:spec2}. This is certainly an over-simplistic interpretation of the line shapes, as some of the features are blends and therefore they will not necessarily have the same width or structure. Typical spreads in our measurement of FWHM are large, on the order of 3000 km s$^{-1}$. Line widths are typically 9000--12,000 km s$^{-1}$.

PS1-10awh, PS1-10ky, and SCP 06F6 show no evidence for decreasing line velocity with time, displaying similar expansion velocities before and after peak (Table \ref{table:spec2} and Figure \ref{radvel}).  This is roughly consistent with the bottom panel of Figure \ref{t_l_r}, where we plot the estimated radii as a function of time, assuming temperatures measured as described in \S \ref{temp} (except in the case of SCP 06F6, where we use the quadratic fit to temperature as a function of time) and luminosities determined from the technique accounting for photometric flux and the red blackbody tail technique in \S \ref{lum}. The estimates of radius are very noisy, but growth with time can be fit with a straight line for these three sources, as would be expected for a roughly constant photospheric velocity. The best-fit expansion velocities of the blackbody photospheres are: $10,600\pm3800$ km s$^{-1}$ for PS1-10awh, $17,700\pm3200$ km s$^{-1}$ for PS1-10ky, $8,600\pm400$ km s$^{-1}$ for SN 2010gx, and $11,400\pm5700$ km s$^{-1}$ for SCP 06F6. Again we remind the reader that the luminosities used to measure these radii are lower limits---especially in the cases of SN 2010gx and SCP 06F6---and this will translate to under-estimates of the expansion velocities. Given the large uncertainties, these velocities measured from radius estimates are consistent with our measurements from the spectral lines. 

The behavior of SN 2010gx is more complex, and comparison of its two phases (when it resembles SCP 06F6 and when it resembles a normal Type Ic SN) is complicated by being forced to use different lines with different systematic uncertainties at different times. 

The early expansion velocity of $\sim$19,000 km s$^{-1}$ observed around Day $-$5 is consistent with the other three sources, but we do not have sufficient spectroscopic information during this SCP 06F6-like phase to discern if it is constant. Later, when SN 2010gx shows \ion{Fe}{2} indicative of a Type Ic SN, its photospheric velocity shows evidence of deceleration from 19,000 km s$^{-1}$ on Day 10 to 15,000 km s$^{-1}$ on Day 30 (also apparent in Figure \ref{spec_multi}). We note that comparisons of velocities measured from different features should be carried out cautiously, as some absorption features may actually be blends of ions that are not properly accounted for in our \verb|SYNOW| modeling, introducing uncertainty into absolute velocity measurements (therefore, it is difficult to state with certainty that the photospheric velocity of SN 2010gx was constant between Day $-$5 and Day 10). However, relative comparisons of velocities measured using the same features should be reliable. Therefore, the data are consistent with constant radial velocity in PS1-10ky, PS1-10awh, and SCP 06F6, and provide evidence for decreasing velocity in the latter SN Type Ic phase of SN 2010gx.


Our analysis for these four ultra-luminous SNe implies that while they are in an SCP 06F6-like phase, showing a relatively featureless continuum with a few broad absorption lines of lighter metals, they do not show any clear sign of change in the rate of photospheric expansion. This apparent lack of deceleration is in direct contrast with SN 2005ap, where the ``W"-shaped feature is observed to decelerate by 4,000 km s$^{-1}$ over seven days in the rest frame \citep{Quimby_etal07}. The apparently constant velocities measured for PS1-10ky, PS1-10awh, and SCP 06F6 are mostly measured from the blue lines of \ion{C}{2}, \ion{Mg}{2}, and \ion{Si}{3}; it is possible that these lines form at a significantly different location in the photosphere from the \ion{O}{2} ``W"-shaped feature, and may therefore show different velocity evolution. SN 2010gx hints that, later in the evolution of these ultra-luminous SNe, the measured velocities will decline, as typically observed during the photospheric phase of SN expansion, while the photosphere recedes into progressively slower-moving ejecta.

\section{Models for SCP 06F6-Like SNe}\label{model}

In this section we consider three physical scenarios that have been proposed to explain ultra-luminous SNe. We compare the predictions of these models with the data presented above, in an effort to define the parameter space that these objects may inhabit. When we discuss bolometric light curves, we are using the luminosities calculated by totaling up the flux in the observed bands and adding on the red tail of a blackbody fit, as described in \S \ref{lum} and plotted in Figure \ref{t_l_r}. These are likely to be slight underestimates of the true bolometric luminosity for PS1-10awh and PS1-10ky, but they are the best estimate we can make given our limited wavelength coverage.

\subsection{Scenario 1: Radioactive Decay}

Free from the H-envelopes associated with most Type II SNe, the photospheric emission from Type I SNe is powered by the radioactive decay of freshly synthesized $^{56}$Ni and $^{56}$Co within the ejecta. The associated optical signal reaches maximum intensity within a month of the explosion at which point the photons diffuse efficiently out of the optically-thick layers causing the light curve to decay.  The characteristic time of photon diffusion, $\tau_c\propto M_{\rm ej}^{3/4}~E_K^{-1/4}$ days, is determined by fundamental ejecta properties \citep{Arnett82}: the total mass ($M_{\rm ej}$; in units of M$_{\odot}$) and the kinetic energy ($E_K$; in units of 10$^{51}$ erg s$^{-1}$).  As originally shown by \citet{Phillips93} for SNe Ia and \citet{Valenti_etal08} and \citet{Drout_etal10} for SNe Ibc, the post-maximum decay rate may serve as a proxy for the light-curve width since broader light-curves are associated with slower decline rates.  We adopt the notation, $\Delta m_{15}$, which represents the magnitude decrease in the 15 days following the peak luminosity.  

The peak luminosity ($L_{\rm peak}$) is primarily determined by the mass of $^{56}$Ni ($M_{\rm Ni}$; see \citealt{Arnett82}). Meanwhile, spectroscopic measurements of the photospheric velocity ($v_{\rm ph}$) directly constrain the quantity $\sqrt{E_K/M_{\rm ej}}$, thus enabling all three physical parameters ($M_{\rm Ni}$, $E_K$, and $M_{\rm ej}$) to be determined uniquely based on three observables: (i) the peak luminosity, (ii) the light-curve width, and (iii) the velocity of the photosphere.

For the rest-frame lightcurve of PS1-10ky, we measure a peak bolometric luminosity of $L_{\rm peak}\approx 3.0 \times 10^{44}$ erg s$^{-1}$ and a post-maximum decay rate of $\Delta m_{15}\approx 0.52$ mag. Based on theoretical models for Type Ic SN light-curves \citep{Valenti_etal08}, this decay rate implies a characteristic time of $\tau_c\approx 14$ days and a radioactive mass of $M_{\rm Ni}\approx 14~\rm M_{\odot}$. With the photospheric velocity of $v_{\rm ph}\approx 18,000~\rm km~s^{-1}$ measured for PS1-10ky, we estimate $M_{\rm ej}\approx 4.4~\rm M_{\sun}$ and $E_K\approx 8.9\times 10^{51}$ erg. This ejecta mass is much less than the nickel mass, implying that a model where radioactive decay powers the bulk of the light curve is unphysical. This is consistent with the findings of \citet{Chatzopoulos_etal09} and \citet{Quimby_etal11}.

At later epochs ($\Delta t\approx 60-300$ days), energy generation within a normal SN's ejecta is dominated by the radioactive decay of $^{56}$Co (half-life $\tau=77$ days). Late-time light-curves can thus directly probe the mass of radioactive decay products, enabling an independent constraint on $M_{\rm Ni}$. To date, there has been no late-time photometry of any SCP 06F6-like source, but the appearance of SN 2010gx as a fairly typical SN Ic at late times implies that such an extended tail should exist to the light curve at a low level. Photometric constraints from 76 days after light curve peak place limits on the $^{56}$Ni mass ejected by SN 2010gx of $\lesssim$ 1 M$_{\odot}$ \citep{Pastorello_etal10}. Late-time observations of PS1-10awh on 29 June 2011 (Day 176) imply a 3$\sigma$ \ips-band limit of $>$23.7 mag, corresponding to a rough limit on the nickel mass of $M_{\rm Ni}\lesssim 13~\rm M_{\odot}$. At z $\approx$ 1, 1 M$_{\odot}$ of radioactive material would translate to an $i$-band magnitude of 26.5 mag at day $\sim$100--150 (rest frame), detectable with deep space-based imaging.

\subsection{Scenario 2: Magnetar Spin-Down}\label{magnetar}

\begin{figure}
\centering
\includegraphics[angle=90, width=9.5cm]{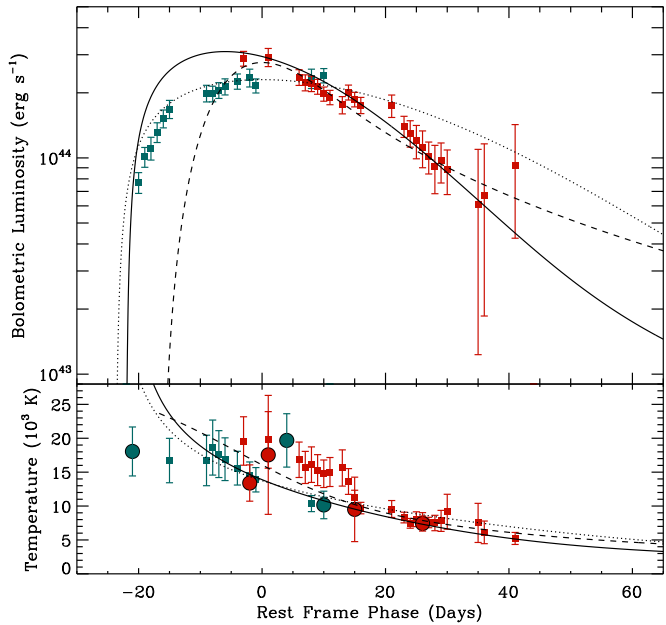}
\caption{Top Panel: Combined bolometric light curve for PS1-10awh and PS1-10ky (colors and symbols as in Fig. \ref{t_l_r}) fit using magnetar spin-down models \citep{Kasen_Bildsten10}. Different lines correspond to fits for different SN ejecta masses: M$_{ej} = 1$ M$_{\odot}$ (dashed), 4 M$_{\odot}$ (solid), and 10  M$_{\odot}$ (dotted); see text for more details. Bottom Panel: temperature evolution plotted as in Fig \ref{t_l_r}, with the effective temperature of the magnetar spin-down models overplotted as lines as in the top panel.}
\label{kb10}
\end{figure}

\citet[][henceforth KB10]{Kasen_Bildsten10} and \citet{Woosley10} have shown that the spin-down of a magnetar can explain the ultra-luminous SNe with broader light curves and slower decays like SN 2005ap \citep{Quimby_etal07} and SN 2008es \citep{Gezari_etal09, Miller_etal09}. However, it remains unclear if they can fit the relatively narrow symmetric light curves of the SCP 06F6-like objects presented here. We use the formalism of KB10 to model SN light curves powered by magnetar spin-down. The magnetar has a rotational energy $E_{p} = 2 \times 10^{50}\ P_{10}^{-2}$ ergs, where $P_{10}$ is its period in units of 10 ms (maximal spin corresponds to $\sim$1 ms). The spin-down timescale due to magnetic dipole radiation is $t_{p} = 1.3\ B_{14}^{-2}\ P_{10}^{2}$ yr, where $B_{14}$ is the magnetar magnetic field strength in units of $10^{14}$ G. It spins down as  $L_{p}(t) = [E_{p} (l - 1)] / [t_{p} (1 + t/t_{p})^{l}]$, where $t$ is the time since birth, $L_{p}$ is the energy the magnetar is depositing into the SN ejecta per unit time, and $l$ is an index that describes the magnetar spin-down ($l = 2$ for magnetic dipole spin-down). The velocity of the SN ejecta is $v_{\rm ej} = [2 (E_{p} + E_{\rm SN}) / M_{\rm ej}]^{1/2}$, where $E_{\rm SN}$ is the SN explosion energy and $M_{\rm ej}$ is the SN ejecta mass. Finally, the diffusion timescale is $t_{d} = [(3 M_{\rm ej} \kappa) / (4 \pi v_{\rm ej} c)]^{1/2}$, where $\kappa$ is the opacity in cm$^{2}$ g$^{-1}$ and $c$ is the speed of light. 

Using these definitions, we solve for the emitted luminosity $L_{e}$ with the differential equation:
\begin{equation}\label{eq:kb10}
\frac{\partial L_{e}(t)}{\partial t}\ =\ {{E_{p}\ (l-1)\ t}\over{t_{d}^{2}\ t_{p}\ (1 + t/t_{p})^{l}}}\ -\ {{t\ L_{e}(t)}\over{t_{d}^{2}}}
\end{equation}
and then multiply the luminosity by a correction factor $(l + 1)/2$ as prescribed by KB10. Using this simple technique, the time to peak luminosity also differs from that solved for in hydrodynamic simulations by KB10. We use their Equation 16 to solve for the ``correct'' peak time, and then scale the time axis of our light curve so that the peak time produced by our equation \ref{eq:kb10} matches it. We find that we can match the published light curves in KB10 well with these simple assumptions.

We leave fixed $l = 2$, $\kappa = 0.2$ cm$^{2}$ g$^{-1}$ (as assumed by KB10), and $E_{\rm SN} = 10^{51}$ erg; we vary $P_{10}$, $B_{14}$, and $M_{\rm ej}$. The bolometric light curve for PS1-10awh and PS1-10ky can be matched reasonably well with $B_{14} = 2.7$, $P_{10} = 0.16$, and $M_{\rm ej} = 4$ M$_{\odot}$, represented as the solid line in Figure \ref{kb10}. We recognize that the light curves for PS1-10awh and PS1-10ky do not match up perfectly, so PS1-10awh is only used as a rough constraint on the rise time. The magnetar in such a model dominates the SN energy with a rotational energy of $8 \times 10^{51}$ erg, and the swept-up shell will expand at 14,900 km s$^{-1}$, also in good accord with measured expansion velocities.

We can also perform a rough comparison with our observed temperature evolution by calculating the effective temperatures of the magnetar models $T = [L_{e} / (4 \pi \sigma v_{\rm ej}^{2} t^{2})]^{1/4}$ where $\sigma$ is the Stefan-Boltzmann constant. We plot the predicted temperature evolution in the bottom panel of Figure \ref{kb10}, and find that it is consistent with our data. 

The light curve can be fit with a range of ejecta masses, from 2--5 M$_{\odot}$. For lower ejecta masses, $P_{10}$ must increase and $B_{14}$ must decrease to fit the data. For example, the best match for $M_{\rm ej} = 3$ M$_{\odot}$ is $B_{14} = 2.5$ and $P_{10} = 0.20$ (implying an expansion velocity of 14,200 km s$^{-1}$), while for $M_{\rm ej} = 5$ M$_{\odot}$ the data can be described well with $B_{14} = 3.4$, $P_{10} = 0.11$ (giving a velocity of 18,800 km s$^{-1}$). In Figure \ref{kb10}, we show two additional models that illustrate why the range of well-fit ejecta masses is limited. The dotted line shows a light curve that matches the peak bolometric luminosity for $M_{\rm ej} = 10$ M$_{\odot}$; clearly the late-time decline of the light curve is too gradual. This lack of good fit for high $M_{\rm ej}$ is due to the maximal spin of a magnetar at $P_{10} = 0.1$ (KB10). There is also no good fit to the light curve for $M_{\rm ej} = 1$ M$_{\odot}$ (dashed line), as a model that produces appropriate peak luminosity and light curve decline also predicts a light curve rise that is far too rapid. We therefore conclude that PS1-10ky and PS1-10awh can be plausibly fit with ejecta masses of $2-5$ M$_{\odot}$,  spanning the commonly observed range of ejecta masses for SNe Type Ic \citep{Drout_etal10}. Different ejecta masses might explain the measured differences in expansion velocity. 

The KB10 magnetar model also makes the distinct prediction that the SN photospheric velocity should be constant with time, consistent with our observations (Table \ref{spec}). This is because the magnetar blows a hot low-density bubble in the center of the SN, producing a dense fast-moving shell that sweeps up the ejecta. Unlike in a normal SN, where the photosphere recedes inward to ejecta with slower velocities, the photosphere stalls at this dense shell in the magnetar model, implying a roughly constant observed expansion velocity. However, it is possible that the magnetar spin-down model would struggle to produce late-time SN Ic emission as seen in SN 2010gx, for this very same reason: the magnetar has cleared out a low-density bubble around it, and the ejecta are no longer expanding homologously but as a relatively thin shell. We also note that the measured evolution of the radius (Figure \ref{t_l_r}) is consistent with a shell that begins expanding at constant velocity from a very small radius around Day $-40$ or Day $-30$, roughly coincident with the first detections of SCP 06F6-like sources; however, our radius calculations are very uncertain, and better temperature determinations before light curve maximum are required to constrain the radius at early times.

\subsection{Scenario 3: Shock Breakout and Circumstellar Interaction}\label{breakout}

The extremely high radiated energies of many ultra-luminous SNe has been explained by interaction with a dense circumstellar medium (CSM). The general idea is that interaction with CSM transforms much of the SN's kinetic energy into radiation before it can be lost to adiabatic expansion \citep[e.g.,][]{Smith_McCray07, Chevalier_Irwin11}.

\citet[][henceforth CI11]{Chevalier_Irwin11} propose that the light curves of SCP 06F6 and related objects might be explained by shock breakout from a dense but truncated CSM. A stellar wind with mass loss rate $\dot{M}$ and velocity $v_{w}$ is expelled from the SN progenitor for a time $t_{\rm ml}$ before the SN explosion, producing a $\rho \propto r^{-2}$ density profile that abruptly drops to zero for radii greater than $r_{w} = t_{\rm ml} v_{w}$. If the diffusion radius of the SN ($r_{d}$) is roughly equivalent to the radius of this stellar wind, SN energy will be transformed into radiation very efficiently, resulting in a luminous burst of radiation with a fairly symmetric light curve. We can use our measured peak bolometric luminosity (L$_{\rm peak} = 3.0 \times 10^{44}$ erg s$^{-1}$) and peak temperature (T$_{\rm peak} = 14,700$ K) to measure this radius (r$_{w} = 3.0 \times 10^{15}$ cm). From the PS1-10ky/PS1-10awh light curve, we estimate a rise time of $t_{d} \approx$ 25 days, which can be roughly equated to the diffusion time $t_{d}$:
\begin{equation}
t_{d}  = 6.6\ k\ D_{\star}\ \textrm{days}
\end{equation}
where $k$ is the opacity in units of 0.34 cm$^{2}$ g$^{-1}$ and D$_{\star}$ parameterizes the mass loss in the stellar wind as $D_{\star} = (\dot{M}/10^{-2}\ M_{\odot})(v_{w}/10\ \rm{km~s}^{-1})$. For an ionized He-rich wind, $k$=0.59 and we find D$_{\star}$ = 6.4. Integrating this density profile out to $r_{w}$, we find that the total mass in the stellar wind is 6.1 M$_{\odot}$. Assuming a Wolf-Rayet (W-R) progenitor star with v$_{w}$ = 1000 km s$^{-1}$ \citep{Nugis_Lamers00}, this corresponds to a wind powered for approximately one year at a mass loss rate $\dot{M} = 6$ M$_{\odot}$ yr$^{-1}$. 

Such a mass loss event, expelling several solar masses in the year before stellar death, is extreme, although such episodes are seen for luminous blue variables \citep[LBVs; e.g.,][]{Davidson_Humphreys97, Smith_Owocki06}. \citet{Pastorello_etal07} discovered such an outburst preceding the type Ib SN 2006jc by two years; they attempt to reconcile the LBV-like outburst with the W-R progenitor of the SN, but point out that no such outburst has ever been observed for a W-R star. They conclude that perhaps the W-R progenitor is in a binary with a LBV, or perhaps such outbursts on W-R stars are exceedingly rare. Meanwhile, \citet{Foley_etal07} point out that a dense He-rich CSM is required to explain the bright \ion{He}{1} emission lines in SNe 1999cq, 2002ao, and 2006jc, and hypothesize that perhaps LBV-like eruptions can persist into the WR stage. However, X-ray measurements of SN 2006jc imply that the CSM may not be particularly massive (of order 0.01 M$_{\odot}$; \citealt{Immler_etal08}), and therefore an episode like that seen in SN 2006jc would need to be significantly scaled up to explain SCP 06F6-like sources.

Given the integrated radiated energy of our combined light curve and $k$ and D$_{\star}$ as described above, we can use Equation 5 from CI11 to find that E$_{51}^2$ / M$_{\rm ej,10}$ = 17, where E$_{51}$ is in units of 10$^{51}$ erg and M$_{\rm ej,10}$ is in units of 10 M$_{\odot}$. For an ejecta mass of 10 M$_{\odot}$, this implies a SN energy of $4.1 \times 10^{51}$ erg. We can also solve for the diffusion radius using Equation 3 from CI11 and find $r_{d} = 1.4 \times 10^{15}$ cm, similar to $r_{w}$ within a factor of two and implying that these objects are in the $r_{w} \approx r_{d}$ regime where radiation is produced most efficiently.

The inner ejecta of a SN explosion typically have a rather flat density profile ($\rho_{\rm in} \propto r^{-1}$ or constant), while the density of the outermost ejecta falls off much more steeply ($\rho_{\rm out} \propto r^{-7}$). The CI11 model assumes that the reverse shock is evolving into the steep outer part of the density profile, but this is only true if $E_{51}^{0.5}\ M_{\rm ej,10}^{-1.5}\ D_{\star}\ t_{\rm SN} < 320$, where $t_{\rm SN}$ is the age of the SN in days. At the beginning of shock breakout, $t_{\rm SN} = t_{d}$ and this condition is marginally satisfied. As shock breakout proceeds, the reverse shock will proceed toward the flatter part of the density profile. Assuming an ejecta mass of 10 M$_{\odot}$, the expanding shell will contain 2.3 M$_{\odot}$ swept up from the circumstellar medium and 1.3 M$_{\odot}$ of reverse-shocked ejecta. This shell will contain $2/3\ E_{51}$, implying a shell velocity of 9,000 km s$^{-1}$, which is lower than we observe. The lower ejecta masses necessary to produce higher velocities imply $E_{51}^{0.5}\ M_{\rm ej,10}^{-1.5}\ D_{\star}\ t_{\rm SN} > 320$ at $t_{d}$, requiring a model for circumstellar interaction that takes into account both the $\rho_{\rm out}$ and $\rho_{\rm in}$ regimes of the ejecta density profile. The development of such a model is outside the scope of this work, but the ejecta velocities measured here (especially the high values measured for SN 2010gx and PS1-10ky) will provide important future constraints.

\citet{Smith_McCray07} propose a scenario similar to CI11 to explain the light curve of SN 2006gy. However, in their model the SN shock does not break out of an r$^{-2}$ smooth wind, but rather plows into an optically-thick shell. The exact density profile of the CSM will not change the light curve significantly, as the light curve is largely determined by simply the radius and mass of the surrounding material. Therefore, the Smith \& McCray model give results similar to those derived above using CI11. 

The same basic physics also operates in the pulsational pair instability model \citep{Woosley_etal07}, favored by \citet{Quimby_etal11} to explain their SCP 06F6-like sources. In this scenario, a massive star has not yet exploded and died, but in approaching the end of its life violently ejects massive shells (some as massive as 18 M$_{\odot}$) that collide with one another and produce luminous outbursts. Some of the theoretical light curves feature rather steep declines reminiscent of the SCP 06F6-like sources. However, the \citet{Woosley_etal07}  model struggles to produce outbursts with velocities significantly in excess of $\sim$5,000 km s$^{-1}$, even for a massive progenitor star of 130 M$_{\odot}$. We therefore find pulsational pair-instability SNe to be an unlikely model for PS1-10awh, PS1-10ky, and SCP 06F6. We also note that such a model does not naturally account for the SN Type Ic features seen in SN 2010gx at late times. 

The shock breakout scenario also predicts that the temperature will rise until luminosity maximum, and then decline after maximum \citep[][CI11]{Ensman_Burrows92}. Our temperature measurements (Figure \ref{t_l_r}) do not clearly follow this trend, appearing to be declining or flat before peak, although these early measurements are extremely noisy. Early detection and deep multi-band photometry will be required in the future to test this prediction of the shock breakout model in additional sources. It is also worth noting that the magnetar model makes a similar prediction for the temperature evolution (KB10), so early-time temperature measurements are unlikely to distinguish between the two models.

We also note that the definition of temperature becomes complex for the non-equilibrium conditions involved in shock breakout. Numerous works \citep[e.g.,][]{Ensman_Burrows92, Katz_etal10, Nakar_Sari10} find that the radiation or color temperature of the shock breakout emission is higher than the effective temperature (defined by the luminosity surface density $L / 4 \pi r^2$). If this holds true in the case of SCP 06F6-like objects, it would increase the wind radius calculated here. The total mass expelled by the progenitor in a wind would increase, but other parameters like $D_{\star}$ and E$_{51}^2$ / M$_{\rm ej,10}$ would not be affected. Such a discrepancy may be necessary if the shock breakout model holds true, as the SN is predicted to expand for a time $t_{d}$ \emph{before} the light curve begins to rise (CI11). However, our measured radii (Figure \ref{t_l_r}) are consistent with explosion at approximately Day $-$30, around the time the light curves begin to rise. A decrease in effective temperature by $\sim$40\% (consistent with that predicted by \citealt{Ensman_Burrows92}) could increase the blackbody radii and provide space for expansion that begins at  Day $-$60 rather that Day $-$30.

Also interesting within the shock breakout picture is the fact that the expansion velocities do not appear to decline over $\sim$30 day baselines (\S \ref{vel}). This might naturally be explained if, indeed, there is no CSM outside of the diffusion radius and therefore the shock is not undergoing any deceleration. However, this is perhaps an oversimplification, as we note that the bolometric light curve for PS1-10awh and PS1-10ky shows some indication of a late-time tail (\S \ref{lum}, Fig \ref{t_l_r}) and we calculate $r_w$ to be slightly larger than $r_d$.  There is therefore the possibility of ongoing CSM interaction at later times, complicating our simple shock breakout picture from a truncated CSM. However, we must remember that these observations are sensitive to an uncertain temperature evolution and they are the product of two independent light curves which have been spliced together.

\subsubsection{Predicted Radio Emission}\label{radio}

We can predict the radio signature for such a shock breakout from a dense wind using standard models of radio SNe \citep{Chevalier96}. While the shock front is inside $r_d$, it propagates as a radiation-dominated shock wave. This shock structure does not give rise to particle acceleration to relativistic energies, so no radio synchrotron emission will result in this early stage. If $r_w > r_d$ and the blast wave continues to interact with CSM after breakout, some particle acceleration might occur, but it would likely result in a relatively short-lived burst of radio emission, as $r_w$ can not be significantly larger than $r_d$ if we are to explain the very high luminosities of SCP 06F6-like objects and the large drop from maximum observed in some objects (e.g., SN 2010gx). In addition, any radio emission will be heavily absorbed due to free-free processes \citep{Weiler_etal02} and synchrotron self absorption \citep{Chevalier98}, so any burst of radio emission will to be faint.

If the stellar wind is truncated as described above, the SN will then expand past the CSM. Even if relativistic particles and magnetic field had been present, their energy densities would rapidly decline due to adiabatic expansion, and the radio flux density would plummet as $S_{\nu} \propto t^{-6}$ \citep{Shklovskii60}. It is possible at this point that a viscous reverse shock develops, but the reverse shock is less powerful than the forward shock, the magnetic field is likely to be weak, and any accelerated particles are subject to loss processes (Coulomb, inverse Compton, and/or synchrotron), so that the reverse shock is unlikely to be a strong source of synchrotron emission. Therefore, after the shock expands past $r_w$, it is unlikely to be a source of radio emission; our EVLA non-detections are not surprising in light of this model.

Finally, we note that we do not expect bright radio emission from the magnetar spin-down scenario (\S \ref{magnetar}). Even Galactic magnetars are typically not detected at radio wavelengths \citep[e.g.,][]{Gaensler_etal01, Burgay_etal06}, and the new-born magnetars proposed by KB10 will additionally suffer free-free absorption from the ionized ejecta in which they are embedded. We therefore do not expect SCP 06F6-like objects to be sources of radio emission, but only the future combination of deep optical time-domain searches with the Large Synoptic Survey Telescope (LSST; \citealt{Ivezic_etal08}) and accompanying blind radio transient surveys with the Australian Square Kilometre Array Pathfinder (ASKAP; \citealt{Johnston_etal07}) will fully test this hypothesis.

\section{Conclusions} \label{conclude}

Using multi-color photometry and multi-epoch spectroscopy, we find that SCP 06F6-like objects radiate $\sim10^{51}$ erg in just a few months, making them amongst the most energetic SNe known. Their peak bolometric magnitudes are typically $\lesssim -$22.5 mag, as compared with $-$19.5 mag for SNe Type Ia \citep{Contardo_etal00}, $-$21.4 mag for the proposed pair-instability SN 2007bi \citep{GalYam_etal09, Young_etal10}, and $-$21.8 mag for the ultra-luminous Type IIn SN 2006gy \citep{Smith_etal07, Ofek_etal07}. Our estimates are consistent with the lower-redshift sources presented in \citet{Quimby_etal11}.

There is no evidence for decreasing radial velocity during the SCP 06F6-like phase in the objects studied here, consistent with the expansion of an optically-thick shell predicted by both the magnetar and circumstellar interaction scenarios. However, the SCP 06F6-like SN 2005ap provides a counter-example to this point, displaying a clear decrease in velocity \citep{Quimby_etal07}. A larger sample of objects with time-series spectroscopy is required to test the prevalence of decelerating photospheric velocities in SCP 06F6-like objects. One source to date, SN 2010gx \citep{Pastorello_etal10} has shown evidence for dissipation of the optically-thick shell and a decrease in optical depth, revealing a normal SN Type Ic in its interior. 

Here, we show that radioactive decay can not explain SCP 06F6-like objects, which rules out pair-instability models \citep[e.g.,][]{Barkat_etal67} for this particular class of ultra-luminous SNe, unlike in the case of the unusually luminous SN 2007bi \citep{GalYam_etal09}. We investigate two physical scenarios that both require an energetic optically-thick shell: the spin-down of a new-born magnetar or shock breakout from a dense circumstellar medium. Both scenarios can fit our data with plausible parameters for a SN Type Ic, although both require rather extreme additional conditions. If magnetars are responsible for SCP 06F6-like objects, they need to be spinning near breakup ($1-2$ ms period). Meanwhile, if shock breakout causes these events, the progenitor must have undergone an LBV-like outburst, expelling several solar masses of H-poor material in just a few years before stellar death; however, no such violent eruption has ever been observed from the Wolf-Rayet stars that are likely the progenitors of these SNe (although dense He-rich CSM has been observed to be present around some SNe Type Ib; \citealt{Pastorello_etal07, Foley_etal07}). One of the most promising strategies for distinguishing between these scenarios in the future is early-time measurement of the photospheric radius, because at the time of light curve rise in the CSM interaction model, the SN will expand from a relatively much larger radius as compared with the magnetar spin-down model. In addition, late-time observations have the potential to distinguish between the two models. For example, the later observations of SN 2010gx may conflict with expectations from the magnetar model, as this scenario predicts that the SN ejecta themselves are being swept into a thin shell by the magnetar wind; however, relatively normal SN Type Ic ejecta are observed in this source at late times.

Any model for SCP 06F6-like sources needs to explain why these ultra-luminous events are only associated with a small fraction of SN explosions; the extreme conditions implied by both models discussed here might justify their rarity, but it remains to be determined if we can expect such fast-spinning magnetars or dense H-poor circumstellar media in sufficient numbers to explain the rates of SCP 06F6-like sources. The lack of detection of host galaxies for PS1-10ky and PS1-10awh (down to $0.02\ L^{\star}$) implies that they are in dwarf galaxies with low metallicities and high specific SFRs. This is a common feature for these type of transients \citep{Quimby_etal11, Pastorello_etal10, Neill_etal11}. As none have been found in an $L^{\star}$-type galaxy, this may point to metallicity playing a key role in the progenitor channel---or it may simply be an artifact of the bulk of cosmic star formation occuring in systems with high specific SFR. A larger sample of SCP 06F6-like sources is required to distinguish between these scenarios.

Moreover, we note that these sources display some degree of homogeneity as a class, showing peak luminosities, expansion velocities, and temperature evolution which agree to better than a factor of two across our (albeit small) sample. Given the widespread cut-off in luminosity blueward of 2500 \AA, we believe SCP 06F6-like sources will be of limited utility for ultraviolet high-redshift absorption-line studies (e.g., Ly $\alpha$), in contrast with the claims of \citet{Quimby_etal11}. As wide-field surveys like PS1 and PTF continue to operate, many more SCP 06F6-like sources will be discovered; PS1 is currently discovering $\sim$1 such source each month at z $\gtrsim$ 0.5. We will be able to test their diversity as a class and measure their rates. Increased sample sizes at a range of redshifts will better constrain the physical properties of these highly energetic explosions, their host stellar populations, and discern between the magnetar spin-down and shock breakout scenarios.

\section{Acknowledgments}

We would like to thank Dan Kasen and Shmuel Balberg. Laura Chomiuk is a Jansky Fellow of the National Radio Astronomy Observatory. Ryan J. Foley is supported by a Clay Fellowship. This discovery was enabled using the PS1 System operated by the PS1 Science Consortium (PS1SC) and its member institutions The PS1 Surveys have been made possible through the combinations of the Institute for Astronomy at the University of Hawaii, The Pan-STARRS Project Office, the Max-Planck Society and its participating institutes, the Max Planck Institute for Astronomy, Heidelberg, and the Max Planck Institute for Extraterrestial Physics, Garching, The Johns Hopkins University, the University of Durham, the University of Edinburgh, the Queen's University of Belfast, the Harvard-Smithsonian Center for Astrophysics, the Las Cumbres Observatory Global Network, and the National Central University of Taiwan. Observations reported here were obtained at the MMT Observatory, a joint facility of the Smithsonian Institution and the University of Arizona. This paper uses data products produced by the OIR Telescope Data Center, supported by the Smithsonian Astrophysical Observatory. The EVLA is run by the National Radio Astronomy Observatory, a facility of the National Science Foundation operated under cooperative agreement by Associated Universities, Inc. Gemini Observatory is operated by the Association of Universities for Research in Astronomy, Inc., under a cooperative agreement  with the NSF on behalf of the Gemini partnership: the National Science Foundation (United States), the Science and Technology Facilities Council (United Kingdom), the  National Research Council (Canada), CONICYT (Chile), the Australian Research Council (Australia), Minist\'{e}rio da Ci\^{e}ncia e Tecnologia (Brazil)  and Ministerio de Ciencia, Tecnolog\'{i}a e Innovaci\'{o}n Productiva (Argentina). We are grateful for access to Gemini under programs GN-2010A-Q-30 and GS-2010B-Q-4 (PI: E.~Berger) and GN-2010B-Q-34 (PI: J.~Tonry). Partial support for this work was provided by National Science Foundation grants AST-1009749 and AST-0807727.

{\it Facilities:} \facility{PS1(GPC1)}, \facility{MMT (Blue Channel Spectrograph, Hectospec)}, \facility{Gemini:Gillett (GMOS)}, \facility{EVLA}, \facility{GALEX}

\bibliography{sl}{}
\LongTables
\clearpage
\begin{deluxetable}{lcccccccccccccccc}
\tablewidth{0 pt}
\tabletypesize{\tiny}
\setlength{\tabcolsep}{0.025in} 
\tablecaption{ \label{table:phot_ky}
 PS1 MDS + GALEX Photometry for PS1-10ky}
\tablehead{ 
\colhead{UT Date} & \colhead{MJD}   & \colhead{Phase}\tablenotemark{a} & \colhead{NUV} & \colhead{\gps} & \colhead{\rps} & \colhead{\ips} & \colhead{\zps}  & \colhead{\yps} \\
\colhead{} & \colhead{}   & \colhead{(days)} & \colhead{(mag)} & \colhead{(mag)} & \colhead{(mag)} & \colhead{(mag)} & \colhead{(mag)}  & \colhead{(mag)}} 
 \startdata
2010-Jun-26 & 55373.5 & $-$12.0  & & & & & & 21.21$\pm$0.18 \\
2010-Jul-11 & 55388.6 & $-$4.3 & &  & & & 21.39$\pm$0.07 & \\
2010-Jul-12 & 55389.5 & $-$3.8 & & 21.47$\pm$0.04 & 21.13$\pm$0.03 & & &  \\
2010-Jul-21 & 55398.6 & 0.8 & & 21.41$\pm$0.12 & 21.10$\pm$0.08 & & &  \\
2010-Jul-23 & 55400.6 & 1.9 & & & & & 21.63$\pm$0.28 &  \\
2010-Jul-25 & 55402.4 & 2.8 & & & & & & 21.30$\pm$0.25 \\
2010-Jul-31 & 55408.5 & 5.9 & $>$23.35 & & & 21.31$\pm$0.07 & & \\
2010-Aug-01 & 55409.5 & 6.4 & $>$23.25 & & & & 21.37$\pm$0.06 & \\
2010-Aug-02 & 55410.5 & 7.0 & & 22.08$\pm$0.13 & 21.30$\pm$0.06 & & & \\
2010-Aug-03 & 55411.5 & 7.5 & & & & 21.48$\pm$0.07 & &  \\
2010-Aug-04 & 55412.5 & 7.9 & & & & & 21.25$\pm$0.07 & \\
2010-Aug-05 & 55413.5 & 8.4 & & 22.05$\pm$0.09 & 21.34$\pm$0.04 & & & \\
2010-Aug-06 & 55414.5 & 9.0 & $>$23.60 & & & 21.46$\pm$0.04 & &  \\
2010-Aug-07 & 55415.6 & 9.5 & & & & & 21.43$\pm$0.10 & \\
2010-Aug-08 & 55416.5 &10.0 & $>$23.58 & 22.40$\pm$0.11 & 21.43$\pm$0.04 & & & \\
2010-Aug-09 & 55417.5 & 10.5 & & & & 21.49$\pm$0.04 & & \\
2010-Aug-11 & 55419.0 & 11.3 & $>$23.58 & & & & & \\
2010-Aug-12 & 55420.5 & 12.0 & $>$23.59 & & & & & \\
2010-Aug-14 & 55422.5 & 13.1 & $>$23.63 & 22.55$\pm$0.22 & 21.51$\pm$0.05 & & & \\
2010-Aug-15 & 55423.5 & 13.6 & & & & 21.63$\pm$0.04 \\
2010-Aug-16 & 55424.5 &14.1 & $>$23.66 & & & & &  \\
2010-Aug-17 & 55425.4 & 14.5 & & 23.03$\pm$0.22 & 21.76$\pm$0.06 & & & \\
2010-Aug-18 & 55426.5 & 15.1 & $>$23.58 & & & 21.64$\pm$0.05 & & \\
2010-Aug-19 & 55427.5 & 15.6 & & & & & 21.66$\pm$0.08 & \\
2010-Aug-20 & 55428.5 & 16.1 & $>$23.54 & 23.33$\pm$0.24 & 22.10$\pm$0.08 & & & \\
2010-Aug-22 & 55430.5 & 17.1 & $>$23.56 & & & & & 21.63$\pm$0.24\\
2010-Aug-28 & 55436.4 & 20.2 & & & & & 21.98$\pm$0.14 & \\
2010-Aug-29 & 55437.5 & 20.7 & & $>$22.68 & 22.05$\pm$0.17 & & & \\
2010-Aug-30 & 55438.5 & 21.2 & & & & 22.19$\pm$0.10 & & \\
2010-Aug-31 & 55439.3 & 21.7 & & & & & 21.77$\pm$0.09 & \\
2010-Sep-02 & 55441.5 & 22.7 & & & & 22.16$\pm$0.07 & & \\
2010-Sep-03 & 55442.5 & 23.3 & & & & & 21.90$\pm$0.11 & \\
2010-Sep-04 & 55443.5 & 23.8 & $>$23.57 & $>$23.82 & 22.72$\pm$0.13 & & & \\
2010-Sep-05 & 55444.4 & 24.3 & & & & 22.43$\pm$0.10 & & \\
2010-Sep-06 & 55445.4 & 24.8 & $>$23.63 & & & & 22.07$\pm$0.13 & \\
2010-Sep-07 & 55446.5 & 25.3 & & $>$23.34 & 22.74$\pm$0.27 & & &  \\
2010-Sep-08 & 55447.4 & 25.8 & & & & 22.33$\pm$0.07 & & \\
2010-Sep-09 & 55448.4 & 26.3 & & & & & 22.17$\pm$0.13 & \\
2010-Sep-10 & 55449.4 & 26.8 & & $>$23.62 & 22.92$\pm$0.15 & & & \\
2010-Sep-11 & 55450.4 & 27.3 & & & & 22.57$\pm$0.22 & & \\
2010-Sep-12 & 55451.5 & 27.9 & & & & & 22.39$\pm$0.26 & \\
2010-Sep-13 & 55452.4 & 28.3 & & $>$23.53 & 23.03$\pm$0.21 & & & \\
2010-Sep-14 & 55453.4 & 28.9 & & & & 22.60$\pm$0.10 & & \\
2010-Sep-15 & 55454.4 & 29.4 & & & & & 22.37$\pm$0.17 & \\
2010-Sep-16 & 55455.4 & 29.9 & & $>$23.43 & & & & \\
2010-Sep-17 & 55456.4 & 30.4 & & & & 22.55$\pm$0.12 & & \\
2010-Sep-18 & 55457.3 & 30.9 & & & & & 22.53$\pm$0.21 & \\
2010-Sep-19 & 55458.3 & 31.4 & & $>$22.02 & $>$22.53 & & & \\
2010-Sep-23 & 55463.4 & 33.5 & & & & & & $>$21.63 \\
2010-Sep-24 & 55463.4 & 34.0 & & & & & & 21.98$\pm$0.32 \\
2010-Sep-25 & 55464.4 & 34.5 & & & & & & $>$21.76 \\
2010-Sep-26 & 55465.5 & 35.0 & & & & 23.01$\pm$0.26 & & \\
2010-Sep-27 & 55466.4 & 35.5 & & & & & 22.89$\pm$0.27 &  \\
2010-Sep-28 & 55467.3 & 36.0 & & $>$23.37 & 23.36$\pm$0.33 & & & \\
2010-Sep-29 & 55468.3 & 36.5 & & & & 23.22$\pm$0.20 & & \\
2010-Oct-04 & 55473.3 & 39.0 & & $>$23.36 & $>$23.76 & & & \\
2010-Oct-06 & 55475.2 & 40.0 & & & & & 22.61$\pm$0.23 & \\
2010-Oct-09 & 55478.2 & 41.5 & & & & & 22.65$\pm$0.23 & \\
2010-Oct-10 & 55479.2 & 42.0 & & $>$23.33 & $>$23.62 & & &  \\
2010-Oct-11 & 55480.2 & 42.6 & & & & $>$23.43 & & \\
2010-Oct-12 & 55481.2 & 43.1 & & & & & $>$22.90 & \\
2010-Oct-13 & 55482.2 & 43.6 & & $>$23.49 & $>$23.55 & & & \\
2010-Oct-14 & 55483.2 & 44.1 & & & & 23.57$\pm$0.25 & &  \\
2010-Oct-15 & 55484.2 & 44.6 & & & & & $>$23.01 &  \\
2010-Oct-17 & 55486.3 & 45.6 & & & & $>$23.11 & & & \\
2010-Oct-23 & 55492.3 & 48.8 & & & & & & $>$21.74 \\
2010-Oct-27 & 55496.3 & 50.8 & & & & & $>$22.14  & \\
2010-Oct-29 & 55498.2 & 51.8 & & & & $>$23.76 & & \\
2010-Oct-30 & 55499.2 & 52.3 & & & & & $>$22.64 & \\
2010-Oct-31 & 55500.2 & 52.8 & & $>$23.40 & $>$23.65 & & & \\
2010-Nov-01 & 55501.4 & 53.4 & & & & $>$23.50 & & \\
2010-Nov-02 & 55502.3 & 53.8 & & & & & $>$22.95 & \\
2010-Nov-03 & 55503.3 & 54.4 & & $>$23.59 & $>$23.59 & & & \\
2010-Nov-07 & 55507.3 & 56.4 & & & & $>$23.52 & & \\
2010-Nov-08 & 55508.2 & 56.9 & & & & & $>$22.57 & \\
2010-Nov-10 & 55510.3 & 57.9 & & & & $>$23.75 & & \\
2010-Nov-12 & 55512.3 & 59.0 & & $>$22.62 & & & & \\
2010-Nov-23 & 55523.3 & 64.6 & & & & & & $>$21.83 \\
2010-Nov-30 & 55530.2 & 68.1 & & & $>$22.73 & & & \\
2010-Dec-05 & 55535.2 & 70.7 & & & & & $>$22.90 &  \\
2010-Dec-06 & 55536.3 & 71.2 & & $>$22.94 & $>$23.57 & & & \\
2010-Dec-17 & 55547.2 & 76.8 & & & & & $>$23.07 & 
\enddata
\tablenotetext{a}{In days relative to peak brightness, corrected for time dilation.}
\end{deluxetable} 
\clearpage

\LongTables
\clearpage
\begin{deluxetable}{lcccccccccccccccc}
\tablewidth{0 pt}
\tabletypesize{\tiny}
\setlength{\tabcolsep}{0.025in} 
\tablecaption{ \label{table:phot_awh}
 PS1 MDS + GALEX Photometry for PS1-10awh}
\tablehead{ 
\colhead{UT Date} & \colhead{MJD}   & \colhead{Phase}\tablenotemark{a} & \colhead{NUV} & \colhead{\gps} & \colhead{\rps} & \colhead{\ips} & \colhead{\zps}  & \colhead{\yps} \\
\colhead{} & \colhead{}   & \colhead{(days)} & \colhead{(mag)} & \colhead{(mag)} & \colhead{(mag)} & \colhead{(mag)} & \colhead{(mag)}  & \colhead{(mag)}} 
 \startdata
2010-Jun-26 & 55373.5 & $-$77.9 & & & & & & $>$21.78 \\
2010-Jul-11 & 55388.6 & $-$69.0 & &   &    &    & $>$23.29 & \\
2010-Jul-12 & 55389.5 & $-$68.5 & & $>$23.61 & $>$23.76 & & & \\
2010-Jul-21 & 55398.6 & $-$63.8 & & & $>$22.62 & & & \\
2010-Jul-23 & 55400.6 & $-$62.7 & & & & & $>$21.43 & \\
2010-Jul-25 & 55402.4 & $-$61.7 & & & & & & $>$21.31\\
2010-Jul-31 & 55408.5 & $-$58.6 & & & & $>$22.94 & & \\
2010-Aug-01 & 55409.5 & $-$58.0 & & &  & & $>$23.23 & \\
2010-Aug-02 & 55410.5 & $-$57.5 & & $>$23.10 & $>$23.03 & & & \\
2010-Aug-03 & 55411.5 & $-$57.0 & & & & $>$23.39 & & \\
2010-Aug-04 & 55412.5 & $-$56.5 & & &  & & $>$23.27 &\\
2010-Aug-05 & 55413.5 & $-$56.0 & & $>$23.47 &  & & & \\
2010-Aug-06 & 55414.5 & $-$55.4 & $>$23.60 & & & $>$23.70 & & \\
2010-Aug-07 & 55415.6 & $-$54.8 & & & & & $>$22.53 & \\
2010-Aug-08 & 55416.5 & $-$54.4 & $>$23.58 & $>$23.60 & $>$23.70 & & &\\
2010-Aug-11 & 55419.0 & $-$53.3 & $>$23.58 & & & & & \\
2010-Aug-12 & 55420.5 & $-$53.9 & $>$23.60 & & & & & \\
2010-Aug-14 & 55422.5 & $-$51.1 & $>$23.63 & $>$23.52 & $>$23.15 & & & \\
2010-Aug-16 & 55424.5 & $-$50.2 & $>$23.66 & & & & $>$23.20 & \\
2010-Aug-17 & 55425.4 & $-$49.7 & & $>$23.48 & $>$23.61 & & & \\
2010-Aug-18 & 55426.5 & $-$49.1 & $>$23.69 & & & $>$23.56 & & \\
2010-Aug-19 & 55427.5 & $-$48.6 & & & & & $>$23.35 & \\
2010-Aug-20 & 55428.5 & $-$48.1 & $>$23.70 & $>$23.43 & $>$23.52 & & & \\
2010-Aug-22 & 55430.5 & $-$47.1 & & & & & & $>$21.85 \\
2010-Aug-28 & 55436.4 & $-$43.9 & & & & & $>$23.04 & \\
2010-Aug-29 & 55437.5 & $-$43.4 & & $>$22.35 & $>$22.65 & & & \\
2010-Aug-30 & 55438.5 & $-$42.8 & & & & $>$23.52 & & \\
2010-Aug-31 & 55439.3 & $-$42.4 & & & & & $>$23.38 & \\
2010-Sep-01 & 55440.5 & $-$41.8 & & $>$23.30 & $>$23.33 & & & \\
2010-Sep-02 & 55441.5 & $-$41.3 & & & & $>$23.57 & & \\
2010-Sep-03 & 55442.5 & $-$40.7 & & & & & $>$22.97 & \\
2010-Sep-04 & 55443.5 & $-$40.2 & & $>$23.63 & $>$23.61 & & & \\
2010-Sep-05 & 55444.4 & $-$39.8 & & & & $>$23.60 & & \\
2010-Sep-06 & 55445.4 & $-$39.2 & & & & & $>$23.27 & \\
2010-Sep-07 & 55446.5 & $-$38.6 & & $>$23.09 & $>$22.95 & & & \\
2010-Sep-08 & 55447.4 & $-$38.2 & & & & $>$23.80 & & \\
2010-Sep-09 & 55448.4 & $-$37.6 & & & & & $>$23.33 & \\
2010-Sep-10 & 55449.4 & $-$37.1 & & $>$23.45 & $>$23.66 & & & \\
2010-Sep-11 & 55450.4 & $-$36.6 & & & & $>$23.19 & & \\
2010-Sep-12 & 55451.5 & $-$36.1 & & & & & $>$22.83 & \\
2010-Sep-13 & 55452.4 & $-$35.6 & & $>$23.64 & $>$23.59 & & & \\
2010-Sep-14 & 55453.4 & $-$35.0 & & & & $>$23.83 & & \\
2010-Sep-15 & 55454.4 & $-$34.5 & & & & & $>$23.29 & \\
2010-Sep-16 & 55455.4 & $-$34.0 & & $>$23.27 & $>$23.56 & & & \\
2010-Sep-17 & 55456.4 & $-$33.5 & & & & $>$23.65 & & \\
2010-Sep-18 & 55457.3 & $-$33.0 & & & & & $>$23.11 & \\
2010-Sep-19 & 55458.3 & $-$32.5 & & $>$21.80 & $>$22.10 & & & \\
2010-Sep-23 & 55463.4 & $-$30.3 & & & & & & $>$21.40 \\
2010-Sep-24 & 55463.4 & $-$29.8 & & & & & & $>$21.87\\
2010-Sep-26 & 55465.5 & $-$28.7 & & & & $>$23.14 & & \\
2010-Sep-27 & 55466.4 & $-$28.2 & & & & & $>$23.18 & \\
2010-Sep-28 & 55467.3 & $-$27.7 & & $>$22.82 & 23.35$\pm$0.35 & & & \\
2010-Sep-29 & 55468.3 & $-$27.2 & & & & 23.40$\pm$0.30 & & \\
2010-Oct-04 & 55473.3 & $-$24.6 & & 22.70$\pm$0.17 & 22.62$\pm$0.14 & & & \\
2010-Oct-06 & 55475.2 & $-$23.6 & & & & & 23.05$\pm$0.33 & \\
2010-Oct-08 & 55477.2 & $-$22.5 & & & & 22.30$\pm$0.10 & & \\
2010-Oct-09 & 55478.2 & $-$22.0 & & & & & 22.55$\pm$0.21 & \\
2010-Oct-10 & 55479.2 & $-$21.5 & & 22.11$\pm$0.12 & 22.04$\pm$0.10 & & & \\
2010-Oct-11 & 55480.2 & $-$21.0 & & & & 22.16$\pm$0.18 & & \\
2010-Oct-12 & 55481.2 & $-$20.5 & & & & & $>$22.60 & \\
2010-Oct-13 & 55482.2 & $-$19.9 & & 21.89$\pm$0.11 & 21.77$\pm$0.07 & & & \\
2010-Oct-14 & 55483.2 & $-$19.4 & & & & 21.94$\pm$0.06 & & \\
2010-Oct-15 & 55484.2 & $-$18.9 & & & & & 21.98$\pm$0.14 & \\
2010-Oct-17 & 55486.3 & $-$17.8 & & & & 21.70$\pm$0.14 & & \\
2010-Oct-27 & 55496.3 & $-$12.6 & & & & & 21.58$\pm$0.23 & \\
2010-Oct-29 & 55498.2 & $-$11.6 & & & & 21.58$\pm$0.06 & & \\
2010-Oct-30 & 55499.2 & $-$11.0 & & & & & 21.66$\pm$0.12 & \\
2010-Oct-31 & 55500.2 & $-$10.5 & & 21.63$\pm$0.07 & 21.43$\pm$0.05 & & & \\
2010-Nov-01 & 55501.4 & $-$9.9 & & & & 21.57$\pm$0.09 & & \\
2010-Nov-02 & 55502.3 & $-$9.4 & & & & & 21.54$\pm$0.11 & \\
2010-Nov-03 & 55503.3 & $-$8.9 & & 21.54$\pm$0.10 & 21.40$\pm$0.07 & & & \\
2010-Nov-07 & 55507.3 & $-$6.8 & & & & 21.32$\pm$0.05 & & \\
2010-Nov-08 & 55508.2 & $-$6.3 & & & & & 21.48$\pm$0.10 & \\
2010-Nov-10 & 55510.3 & $-$5.3 & & & & 21.21$\pm$0.04 & & \\
2010-Nov-12 & 55512.3 & $-$4.2 & & 21.54$\pm$0.23 & & & & \\
2010-Nov-23 & 55523.3 & 1.6 & & & & & & $>$21.61 \\
2010-Dec-04 & 55534.3 & 7.3 & & & & 21.28$\pm$0.08 & & \\
2010-Dec-05 & 55535.2 & 7.8 & & & & & 21.23$\pm$0.08 & \\
2010-Dec-06 & 55536.3 & 8.4 & & $>$22.21 & & & &\\
2010-Dec-17 & 55547.2 & 14.2 & & & & & 21.64$\pm$0.10 &
\enddata
\tablenotetext{a}{In days relative to peak brightness, corrected for time dilation.}
\end{deluxetable} 
\clearpage

\end{document}